\documentclass[twocolumn,showpacs,preprintnumbers,amsmath,amssymb,superscriptaddress]{revtex4-1}

\usepackage{graphicx}
\usepackage{dcolumn}
\usepackage{bm}
\usepackage[]{units}
\usepackage{hyperref}
\usepackage{dcolumn}
\usepackage{multirow}
\usepackage{gensymb}
\usepackage{lineno}
\begin{document}


\title{Measurements using the inelasticity distribution of multi-TeV neutrino interactions in IceCube}

\affiliation{III. Physikalisches Institut, RWTH Aachen University, D-52056 Aachen, Germany}
\affiliation{Department of Physics, University of Adelaide, Adelaide, 5005, Australia}
\affiliation{Dept.~of Physics and Astronomy, University of Alaska Anchorage, 3211 Providence Dr., Anchorage, AK 99508, USA}
\affiliation{Dept.~of Physics, University of Texas at Arlington, 502 Yates St., Science Hall Rm 108, Box 19059, Arlington, TX 76019, USA}
\affiliation{CTSPS, Clark-Atlanta University, Atlanta, GA 30314, USA}
\affiliation{School of Physics and Center for Relativistic Astrophysics, Georgia Institute of Technology, Atlanta, GA 30332, USA}
\affiliation{Dept.~of Physics, Southern University, Baton Rouge, LA 70813, USA}
\affiliation{Dept.~of Physics, University of California, Berkeley, CA 94720, USA}
\affiliation{Lawrence Berkeley National Laboratory, Berkeley, CA 94720, USA}
\affiliation{Institut f\"ur Physik, Humboldt-Universit\"at zu Berlin, D-12489 Berlin, Germany}
\affiliation{Fakult\"at f\"ur Physik \& Astronomie, Ruhr-Universit\"at Bochum, D-44780 Bochum, Germany}
\affiliation{Universit\'e Libre de Bruxelles, Science Faculty CP230, B-1050 Brussels, Belgium}
\affiliation{Vrije Universiteit Brussel (VUB), Dienst ELEM, B-1050 Brussels, Belgium}
\affiliation{Dept.~of Physics, Massachusetts Institute of Technology, Cambridge, MA 02139, USA}
\affiliation{Dept. of Physics and Institute for Global Prominent Research, Chiba University, Chiba 263-8522, Japan}
\affiliation{Dept.~of Physics and Astronomy, University of Canterbury, Private Bag 4800, Christchurch, New Zealand}
\affiliation{Dept.~of Physics, University of Maryland, College Park, MD 20742, USA}
\affiliation{Dept.~of Physics and Center for Cosmology and Astro-Particle Physics, Ohio State University, Columbus, OH 43210, USA}
\affiliation{Dept.~of Astronomy, Ohio State University, Columbus, OH 43210, USA}
\affiliation{Niels Bohr Institute, University of Copenhagen, DK-2100 Copenhagen, Denmark}
\affiliation{Dept.~of Physics, TU Dortmund University, D-44221 Dortmund, Germany}
\affiliation{Dept.~of Physics and Astronomy, Michigan State University, East Lansing, MI 48824, USA}
\affiliation{Dept.~of Physics, University of Alberta, Edmonton, Alberta, Canada T6G 2E1}
\affiliation{Erlangen Centre for Astroparticle Physics, Friedrich-Alexander-Universit\"at Erlangen-N\"urnberg, D-91058 Erlangen, Germany}
\affiliation{D\'epartement de physique nucl\'eaire et corpusculaire, Universit\'e de Gen\`eve, CH-1211 Gen\`eve, Switzerland}
\affiliation{Dept.~of Physics and Astronomy, University of Gent, B-9000 Gent, Belgium}
\affiliation{Dept.~of Physics and Astronomy, University of California, Irvine, CA 92697, USA}
\affiliation{Dept.~of Physics and Astronomy, University of Kansas, Lawrence, KS 66045, USA}
\affiliation{SNOLAB, 1039 Regional Road 24, Creighton Mine 9, Lively, ON, Canada P3Y 1N2}
\affiliation{Department of Physics and Astronomy, UCLA, Los Angeles, CA 90095, USA}
\affiliation{Dept.~of Astronomy, University of Wisconsin, Madison, WI 53706, USA}
\affiliation{Dept.~of Physics and Wisconsin IceCube Particle Astrophysics Center, University of Wisconsin, Madison, WI 53706, USA}
\affiliation{Institute of Physics, University of Mainz, Staudinger Weg 7, D-55099 Mainz, Germany}
\affiliation{Department of Physics, Marquette University, Milwaukee, WI, 53201, USA}
\affiliation{Physik-department, Technische Universit\"at M\"unchen, D-85748 Garching, Germany}
\affiliation{Institut f\"ur Kernphysik, Westf\"alische Wilhelms-Universit\"at M\"unster, D-48149 M\"unster, Germany}
\affiliation{Bartol Research Institute and Dept.~of Physics and Astronomy, University of Delaware, Newark, DE 19716, USA}
\affiliation{Dept.~of Physics, Yale University, New Haven, CT 06520, USA}
\affiliation{Dept.~of Physics, University of Oxford, 1 Keble Road, Oxford OX1 3NP, UK}
\affiliation{Dept.~of Physics, Drexel University, 3141 Chestnut Street, Philadelphia, PA 19104, USA}
\affiliation{Physics Department, South Dakota School of Mines and Technology, Rapid City, SD 57701, USA}
\affiliation{Dept.~of Physics, University of Wisconsin, River Falls, WI 54022, USA}
\affiliation{Dept.~of Physics and Astronomy, University of Rochester, Rochester, NY 14627, USA}
\affiliation{Oskar Klein Centre and Dept.~of Physics, Stockholm University, SE-10691 Stockholm, Sweden}
\affiliation{Dept.~of Physics and Astronomy, Stony Brook University, Stony Brook, NY 11794-3800, USA}
\affiliation{Dept.~of Physics, Sungkyunkwan University, Suwon 440-746, Korea}
\affiliation{Dept.~of Physics and Astronomy, University of Alabama, Tuscaloosa, AL 35487, USA}
\affiliation{Dept.~of Astronomy and Astrophysics, Pennsylvania State University, University Park, PA 16802, USA}
\affiliation{Dept.~of Physics, Pennsylvania State University, University Park, PA 16802, USA}
\affiliation{Dept.~of Physics and Astronomy, Uppsala University, Box 516, S-75120 Uppsala, Sweden}
\affiliation{Dept.~of Physics, University of Wuppertal, D-42119 Wuppertal, Germany}
\affiliation{DESY, D-15738 Zeuthen, Germany}

\author{M.~G.~Aartsen}
\affiliation{Dept.~of Physics and Astronomy, University of Canterbury, Private Bag 4800, Christchurch, New Zealand}
\author{M.~Ackermann}
\affiliation{DESY, D-15738 Zeuthen, Germany}
\author{J.~Adams}
\affiliation{Dept.~of Physics and Astronomy, University of Canterbury, Private Bag 4800, Christchurch, New Zealand}
\author{J.~A.~Aguilar}
\affiliation{Universit\'e Libre de Bruxelles, Science Faculty CP230, B-1050 Brussels, Belgium}
\author{M.~Ahlers}
\affiliation{Niels Bohr Institute, University of Copenhagen, DK-2100 Copenhagen, Denmark}
\author{M.~Ahrens}
\affiliation{Oskar Klein Centre and Dept.~of Physics, Stockholm University, SE-10691 Stockholm, Sweden}
\author{I.~Al~Samarai}
\affiliation{D\'epartement de physique nucl\'eaire et corpusculaire, Universit\'e de Gen\`eve, CH-1211 Gen\`eve, Switzerland}
\author{D.~Altmann}
\affiliation{Erlangen Centre for Astroparticle Physics, Friedrich-Alexander-Universit\"at Erlangen-N\"urnberg, D-91058 Erlangen, Germany}
\author{K.~Andeen}
\affiliation{Department of Physics, Marquette University, Milwaukee, WI, 53201, USA}
\author{T.~Anderson}
\affiliation{Dept.~of Physics, Pennsylvania State University, University Park, PA 16802, USA}
% ...

\date{\today}

\collaboration{IceCube Collaboration}
\email{Email: analysis@icecube.wisc.edu}
\noaffiliation

\begin{abstract}

Inelasticity--the fraction of a neutrino's energy transferred to hadrons--is a quantity of interest in the study of astrophysical and atmospheric neutrino interactions at multi-TeV energies with IceCube.  In this work, a sample of contained neutrino interactions in IceCube is obtained from 5 years of data and classified as 2650 tracks and 965 cascades.  Tracks arise predominantly from charged-current $\nu_{\mu}$ interactions, and we demonstrate that we can reconstruct their energy and inelasticity.  The inelasticity distribution is found to be consistent with the calculation of Cooper-Sarkar {\it et al.} across the energy range from $\sim$ 1 TeV to $\sim$ 100 TeV.  Along with cascades from neutrinos of all flavors, we also perform a fit over the energy, zenith angle, and inelasticity distribution to characterize the flux of astrophysical and atmospheric neutrinos.  The energy spectrum of diffuse astrophysical neutrinos is well-described by a power-law in both track and cascade samples, and a best-fit index $\gamma=2.62\pm0.07$ is found in the energy range from 3.5 TeV to 2.6 PeV.  Limits are set on the astrophysical flavor composition that are compatible with a ratio of $\left(\frac{1}{3}:\frac{1}{3}:\frac{1}{3}\right)_{\oplus}$.  Exploiting the distinct inelasticity distribution of $\nu_{\mu}$ and $\bar{\nu}_{\mu}$ interactions, the atmospheric $\nu_{\mu}$ to $\bar{\nu}_{\mu}$ flux ratio in the energy range from 770 GeV to 21 TeV is found to be $0.77^{+0.44}_{-0.25}$ times the calculation by Honda {\it et al.} Lastly, the inelasticity distribution is also sensitive to neutrino charged-current charm production.  The data are consistent with a leading-order calculation, with zero charm production excluded at $91\%$ confidence level.  Future analyses of inelasticity distributions may probe new physics that affects neutrino interactions both in and beyond the Standard Model.

\end{abstract}
\maketitle

\section{Introduction}

The observation of astrophysical neutrinos \cite{Aartsen:2013jdh,Aartsen:2014gkd} was a landmark in high-energy astrophysics.  It introduced a new probe that is directionally sensitive to high-energy hadronic particle accelerators in the universe.  Neutrinos provide both good directional information, unaffected by magnetic fields, and extremely long range, allowing us to probe accelerators at cosmologically significant distances.   Measurements of the flux and energy spectrum \cite{Aartsen:2015zva,Aartsen:2017mau} and flavor composition \cite{Aartsen:2015ivb,Aartsen:2015knd} are so far fully compatible with conventional acceleration models, but more exotic production mechanisms cannot be ruled out.

 At the same time, the observation of high-energy astrophysical and atmospheric neutrinos by detectors like IceCube has opened up the study of neutrino interactions at energies orders of magnitude above those accessible at terrestrial accelerators.  Already, the 1 km$^3$ IceCube neutrino observatory has used atmospheric and astrophysical neutrinos to measure neutrino absorption in the Earth, and from that determined the neutrino-nucleon cross section at energies from 6.3 TeV to 980 TeV to be in agreeement with the Standard Model prediction \cite{Aartsen:2017kpd}.
 
In this paper, we report on a new study of high-energy charged-current (CC) $\nu_\mu$ interactions contained within IceCube's instrumented volume.  These interactions produce a cascade of hadrons and a muon, an event topology known as a starting track.  By estimating the hadronic cascade and muon energies separately, we can estimate the inelasticity of each interaction -- the ratio of hadronic cascade energy to total neutrino energy \cite{Dissertation}.    The central 90\% of neutrinos have estimated energies in the range from 1.1 TeV to 38 TeV,  energies far beyond the reach of terrestrial accelerators.  For example, the NuTeV data were used to measure inelasticity distributions at energies up to 250 GeV \cite{Schienbein:2007fs}, while earlier experiments were limited to lower energies \cite{Aubert:1974en}.  

The starting track data, together with a similarly obtained set of cascades due to all neutrino flavors are binned by reconstructed energy and zenith angle and (for the tracks) inelasticity.  This data is fitted to a neutrino flux model containing both atmospheric and astrophysical neutrinos.   From this, we present measurements of neutrino inelasticity and estimate the fraction of neutrino interactions that produce charmed particles.  We also compare the track and cascade samples, study whether they have the same astrophysical neutrino flux spectral indices, and constrain the flavor composition of astrophysical neutrinos.   Finally, we measure the ratio of neutrinos to antineutrinos in  the atmospheric neutrino flux.

\section{$\nu$ interactions and inelasticity}

The inelasticity distribution of neutrinos is expected to be well described by the Standard Model for weak interactions.  At TeV energies, the interactions are dominated by Deep Inelastic Scattering (DIS).  Neutrinos interact with quarks in nuclear targets (hydrogen and oxygen) via charged-current (CC) and neutral-current (NC)  reactions.  In NC interactions, the neutrino interacts via $Z^0$ exchange, leaving the quark flavor unchanged.  In CC $\nu$ interactions via $W^{\pm}$ exchange, the quark charge changes by one, turning a charge $+2/3$ quark into a charge $-1/3$ one, while for $\overline\nu$, the reverse reactions occur, but with a different inelasticity distribution.  $W^\pm$ also interact with sea quarks and antiquarks in the nucleus in similar charge-changing reactions, so the differences between neutrinos and antineutrinos largely disappear at very high energies.    The relative probability for producing a given final state quark depends on the Cabibbo-Kobayashi-Maskawa  (CKM) matrix elements, so reactions involving $u\rightarrow d$ and $s\rightarrow c$ predominate.   

The neutrino-nucleon cross section involves four main kinematic variables:  $s$ is the neutrino-nucleon center of mass energy squared, $Q^2$ is the negative of the 4-momentum transfer squared, Bjorken$-x$, the fraction of the struck nucleon's momentum carried by the struck quark, and the inelasticity, $y$.   These quantities are related through $x=Q^2/2ME_{\nu}y$, where $M$ is the nucleon mass and $E_{\nu}$ is the incoming neutrino energy.   Most of the interactions discussed in this paper involve struck partons with $10^{-2} < x < 10^{-1}$, and the typical $Q^2$ is $\sim M_{W,Z}^2\approx 6\times10^3$ GeV$^2$.  

 The double-differential cross section for a neutrino with energy $E_\nu$ is
\begin{multline}
\frac{d\sigma^{\nu,\overline\nu}}{dxdy} =  \frac{G_F^2ME_\nu}{2\pi} \left(\frac{M_V^2}{M_V^2+Q^2}\right)^2 \big[(1+(1-y)^2)F_2(x,Q^2)\\-y^2F_L(x,Q^2)\pm (1-(1-y)^2)F_3(x,Q^2)\big]
 \label{eq:sigma}
\end{multline}
where the $+$ sign is for neutrinos, and the $-$ is for antineutrinos.  $M_V$ is the vector boson mass ($M_W$ for charged current interactions and $M_Z$ for neutral current interactions) and $G_F$ is the Fermi coupling constant \cite{DIS}.  The cross sections depend on three nucleon structure functions: $F_2$, $F_3$, and $F_L$.  

Assuming standard vector minus axial vector ($V-A$) coupling and an isoscalar target, at leading order these structure functions are related to the quark parton distribution functions (PDFs) of the target nucleon, $q_i(x,Q^2)$, according to \cite{DIS} 
\begin{align}
F_2(x,Q^2) &= 2x\Sigma_i (q_i(x,Q^2) + \overline q_i(x,Q^2)), \\
F_3(x,Q^2) &= 2x \Sigma_i (q_i(x,Q^2) - \overline q_i(x,Q^2)), \\
F_L (x,Q^2)&= 0.
\end{align}
The longitudinal structure function $F_L$ becomes non-zero in higher order calculations.  It should be noted that measurements of the nuclear structure function $F_2(x,Q^2)$ inferred from neutrino-DIS on an iron target may be slightly different from those observed in charged lepton DIS at low $Q^2$ \cite{Kalantarians:2017mkj}.

This paper uses as a baseline a next-to-leading order calculation (`CSMS') which uses the DGLAP formalism for parton evolution \cite{CooperSarkar:2011pa}.   The calculation uses the HERAPDF1.5 \cite{Radescu2013}  parton distribution functions, with the MSTW2008 and CT10 \cite{Lai:2010vv} distributions used as a standard for comparison.  In the relevant energy range, the calculation has an expected uncertainty of a few percent and is consistent with an independent calculation that used the MSTW2008 PDFs \cite{Connolly:2011vc}.

Equation \ref{eq:sigma} does not account for threshold behaviors which are important for heavy quark production.  For this analysis, charm production is most important.  In the relevant energy range, charm production is about 10\% of the total production.   Because of the quark mass, heavy quark production near its production threshold occurs at a larger average inelasticity than for light quarks at the same energy \cite{Barge:2016uzn}.  In the case of charm quarks, the inelasticity also tends to be higher even at energies far above the production threshold since they originate primarily from scattering off sea $s$-quarks.  In addition, heavy quarks may decay semileptonically, transferring some of their energy to a muon. This muon will not be separately visible in IceCube.  Instead, it will increase the apparent brightness of the primary muon, leading to a higher measured muon energy and lower measured inelasticity.  However, due to the small $\sim 10\%$ branching ratio of muonic charm decays, IceCube is primarily sensitive to the signature of larger inelasticity rather than the di-muon signal.

The CSMS calculation is for neutrino DIS on a single nucleon target; there are a few other contributions to consider.   Water contains hydrogen, so is not a perfectly isoscalar target.  The MINERvA collaboration has seen that nuclear shadowing reduces the cross section for neutrino-heavy ion interactions with $x<0.1$ \cite{Mousseau:2016snl}.  The suppression is expected to increase with decreasing $x$ values \cite{Schienbein:2009kk}.  However, oxygen is a small nucleus, and we expect the cross section reduction to be small for the moderate $x $ values probed here. 

An additional contribution to the cross section is due to neutrino electromagnetic (diffractive) interactions with the Coulomb field of the nuclei, but this is small for low-$Z$ nuclei like hydrogen and oxygen \cite{Seckel:1997kk,Alikhanov:2014uja}.   These effects are not expected to be significant for this analysis. 

 \section{Neutrino Sources and Simulation}
 \label{sec:sources}

This analysis uses atmospheric and astrophysical neutrinos as sources.   The signals observed in IceCube depend on the neutrino fluxes incident on the Earth, their absorption in the Earth, their interactions in and (for backgrounds) around IceCube, their propagation through the detector and the detector response.  We will briefly discuss these factors, with a special emphasis on the interactions, which determine the inelasticity.   

Since most of the results in this paper are based on comparisons of data with various simulations, this section will focus on the physics models used in the simulations.  These models are implicit in the results presented, and the systematic errors depend on the assumptions used in the simulations;  these uncertainties will be discussed when the individual results are presented. 

\subsection{Neutrino Sources}

Conventional atmospheric neutrinos come from pions and kaons that are produced in cosmic-ray air showers.  At the energies relevant for this analysis, the flux roughly follows a power-law spectrum $dN/dE_\nu \propto E^{-(\gamma +1)}$, where $\gamma$ is the spectral index for the cosmic-ray energy spectrum.  Below the cosmic-ray knee, $\gamma\approx 2.7$, while at higher energies $\gamma\approx 3.0$.  The flux is highest for near-horizontal incidence.  This analysis uses the HKKMS \cite{Honda:2006qj} flux calculations extrapolated upward in energy, with a modification to account for the knee of the cosmic-ray spectrum \cite{Aartsen:2016xlq}.  This calculation is in good agreement with previous IceCube measurements  \cite{Aartsen:2012uu,Aartsen:2015xup,Aartsen:2017nbu}.  

Prompt atmospheric neutrinos come from the decay of charmed mesons produced in cosmic-ray air showers. They have yet to be observed but are expected to have a hard spectrum that follows that of the primary cosmic-rays. This analysis uses as a baseline the BERSS \cite{Bhattacharya:2015jpa} perturbative QCD calculation of the prompt flux, which is tied to recent data from RHIC and the LHC, and consistent with similar independent calculations \cite{Garzelli2015,Gauld2016}.

The number of observed prompt and atmospheric neutrinos requires an important adjustment to account for the IceCube `self-veto' - a downward-going atmospheric neutrino will be accompanied by an air shower and muon bundle, which may overshadow the neutrino and cause the event to fail the event containment cuts, and so not register as a starting event.  This probability for a self-veto depends on whether the neutrino is prompt or conventional and on the neutrino energy and zenith angle.  This analysis uses the probabilities calculated in Ref.~\cite{Gaisser:2014bja}.  The muon threshold is taken to be 100 GeV; this is the minimum muon energy which is likely to trigger the self-veto.   The appropriateness of the 100 GeV threshold was verified using detector simulations using the CORSIKA program \cite{CORSIKA}. 

Previous IceCube measurements have found that, in this energy range, the astrophysical flux is consistent with a single power law.  Our fit is based on this single power law. 

The neutrinos observed in IceCube may pass through the Earth before reaching the detector; the absorption in the Earth is simulated following the Standard Model cross sections \cite{CooperSarkar:2011pa}.  The Earth's density profile is assumed to follow the Preliminary Reference Earth Model (PREM) \cite{PREM}.

\subsection{Neutrino Interactions and Cherenkov light emission}
 
 Neutrino interactions in and around the detector are modeled following the CSMS calculation, as described in the previous section.   For NC interactions, the inelasticity is the fraction of the neutrino energy that is transferred to the struck nucleus; the remaining energy escapes from the detector.    For CC interactions, the remainder of the energy is transferred to a charged lepton.   IceCube observes the Cherenkov light emitted by the lepton and its secondary relativistic charged particles.  The hadronic cascade from the struck nucleus also produces Cherenkov light.  Each type of lepton produces light very differently in the detector. 
 
 Electrons produce electromagnetic cascades; the light output is proportional to the electron energy, with the relationship determined from detailed simulations  \cite{Radel:2012ij}.  At low energies, they are treated as point sources, while at energies above 1 TeV, the longitudinal profile of a cascade is approximated using a sequence of uniformly spaced point sources.  At higher energies above 1 PeV, the longitudinal profile includes the LPM effect \cite{Klein:1998du}.
 
 As they traverse the detector, muons radiate energy via ionization, bremsstrahlung, direct pair production and photonuclear interactions; these are modeled following Refs. \cite{Chirkin:2004hz,Koehne:2013gpa}.
 
Tau leptons are propagated through the detector in a manner similar to muons, with adjustments for their higher mass.  IceCube simulations allow the taus to decay into $\overline\nu_\tau e\nu_e$, $\overline\nu_\tau \mu\nu_\mu$ or $\overline\nu_\tau$ plus hadrons.   For the leptonic decays, the leptons are propagated through the detector starting at the $\tau$ decay point, while the neutrinos escape.  For the hadronic decays, the hadronic energy is summed, producing another hadronic cascade, at the point where the $\tau$ decays.   Because of the energy carried off by the escaping neutrinos, $\nu_\tau$ will deposit less energy in the detector and appear similar to lower energy $\nu_e$ and $\nu_\mu$.  In muonic decays, the outgoing muon has a lower average energy than in corresponding $\nu_{\mu}$ interactions.  This can affect the measured inelasticity distribution \cite{DeYoung2007}.

This analysis is sensitive to charm production in neutrino interactions.  DIS interactions produce charm quarks, which then hadronize, forming charmed hadrons and baryons, with lifetimes of order $10^{-12}$ s.  We use the hadronization fractions into $D^{+}$, ${D^0}$, $D_s^{+}$, and $\Lambda_c^{+}$ from the calculation in Ref.~\cite{Lellis:2002} at 100 GeV, which is also used in the GENIE event generator \cite{genie}.  If the charmed hadrons have energies above about 10 TeV, they may interact in the ice and lose energy before they decay.  The interaction probability depends on the individual hadron-ice cross sections and hadron lifetimes. This analysis used parametrizations of the charm cross sections in CORSIKA \cite{CORSIKA}.  Since the parametrizations are only valid above 1 PeV, we extrapolate downward in energy using the kaon-nucleon and nucleon-nucleon cross sections scaled by 88\% and 122\%, respectively, to match CORSIKA parametrization at 1 PeV for charm mesons and baryons.  This leads to critical energies for the $D^+$, $D^0$, $D_s^+$ and $\Lambda_c^+$ of 22 TeV, 53 TeV, 47 TeV and 104 TeV, respectively.  When the charmed hadrons interact, they lose energy.  We use the approach of Ref.~\cite{charm} to parametrize their energy loss distribution.  The observable effect of multiple charm interactions is the production of a low energy muon after a semileptonic decay, mimicking a high inelasticity track for $\nu_e$ or $\nu_\tau$ interactions.  However, due to the 10\% muonic branching ratio, this is not a large effect, and the approximate treatment of charm interactions here does not significantly influence later results on charm production.

The conversion between hadronic cascade energy and light follows Ref.~\cite{Aartsen:2013vja}.   Hadronic cascades produce less light than electromagnetic cascades of the same energy, with larger cascade-to-cascade variation.  At 100 TeV, a hadronic cascade produces an average of 89\% of the light of an equivalent energy electromagnetic cascade.    The difference drops with increasing energy.  Since electromagnetic and hadronic cascades cannot be readily distinguished in IceCube, we will refer to the visible cascade energy as the energy of an electromagnetic cascade producing an equivalent amount of light as the hadronic cascade.

\subsection{Optical transmission through Antarctic ice}

The emitted Cherenkov light travels through the ice, where it may scatter or be absorbed before reaching an optical sensor.   Because the sensor array is sparse, only a tiny fraction of the produced Cherenkov photons are observed, and they are likely to scatter multiple times before reaching a sensor.  Because of this, the signals seen by IceCube are sensitive to the optical properties of the ice.  The scattering and absorption lengths in the ice depend on the position in the ice (largely, but not entirely on the depth below the surface) and on the direction of photon propagation   \cite{Aartsen:2013rt,Aartsen:2013ola}.   These optical properties have been measured using a variety of means, including laser and light emitting diode (LED) signals and cosmic-ray muons.   The optical properties of the ice vary strongly with depth, with certain depth ranges containing ``dust layers'' with very large absorption and scattering, and other depths providing good transmission.  This positional dependence is accounted for with an ice model.  

This analysis used as a baseline the ``SPICE Mie" ice model \cite{Aartsen:2013ola}.  It is based on measurements of the ice properties using LEDs mounted in the detector housings, supplemented with parameterizations to give the wavelength dependence of the scattering and absorption lengths.  It divides the ice into 10 m thick layers and determines the scattering and absorption lengths separately for each depth range.  It does not account for tilts of the dust layers ({\it i.e.} variation of  optical properties depending on horizontal position), or for anisotropy in the ice.

Near the optical sensors, there is additional scattering in the `hole ice' -- the refrozen column of melted water created by the drill during deployment.  This ice contains a central column of bubbles \cite{Rongen:2016sbk}.  In our baseline simulations, it is treated as having a scattering length of 50 cm.

\section{Detector and Data}

The Cherenkov light is detected with an array of 5,160 digital optical modules (DOMs), spread over 1 km$^3$ \cite{Halzen:2010yj,Aartsen:2016nxy}.  The DOMs are deployed in 86 vertical strings, each containing 60 DOMs.  Seventy-eight of the strings are laid out on a 125 m triangular grid.  On those strings, the DOMs are deployed with 17 m vertical spacing between 1450 and 2450 m below the surface.  The remaining strings are laid out in the middle of the array, with smaller string-to-string spacing \cite{Collaboration:2011ym}.  On those strings, the DOMs are deployed closer together at depths between about 2,000 and 2450 m.  

Each DOM collects data independently, sending digitized data to the surface  \cite{Achterberg:2006md,Abbasi:2008aa}.   The optical sensor is a 10 inch photomultiplier tube (PMT) \cite{Abbasi:2010vc}.  The PMT is read out with a data-acquisition system comprising two waveform digitizer systems which are triggered by a discriminator with a threshold of about 1/4 of a typical photoelectron pulse height.  One records data with 14-bit dynamic range at 300 megasamples/s for 400 nsec, and the other collects data with 10-bit dynamic range at 40 megasamples/s for 6.4 $\mu$s.  

All of the digitized data is sent to the surface where a trigger system monitors the incoming data and creates an event when certain conditions are satisfied.  For this analysis, the main trigger required 8 hits within a sliding 5 $\mu$s window.  When a trigger occurs, all of the data within  4 $\mu$s before or 6 $\mu$s after the trigger time is saved and sent to an on-line computer farm for further processing.

The farm applies a number of different selection algorithms to each event.  Each algorithm tests for different classes of interesting events, albeit with significant overlap.  This analysis used as input all events that passed either the cascade or muon track filters.  These filters have very loose cuts.  Simulation studies show that they capture more than 99.5\% of the events that pass the other cuts that are applied here. 

This analysis uses data collected between May 2011 and May 2016, a total live time of 1734 days during which the detector was in its complete 86-string configuration.  During the design of the analysis, 10\% of the data was used for testing and the remaining 90\% was kept blind.

\section{Event Selection}

The analysis aims to select a high-purity sample of neutrino interactions that occur within the detector.  Starting tracks are of the greatest interest here, but cascades are more numerous for an astrophysical flux with equal flavor ratio and provide important constraints for many of the fits discussed here.   For both channels, the first step of the analysis is to select a clean sample of starting events.  The initial cut selects events with more than 100 observed photoelectrons (PE), which captures most contained neutrino interactions above $1\;\rm{TeV}$. 

The next selection applies an outer layer veto to reject events that come from charged particles that enter from outside the detector.  The veto uses DOMs on the top, bottom and sides of the detector.  It also includes DOMs in a horizontal layer that passes through the detector, below the ``dust layer'' that allows charged particles to sneak in undetected.  The veto is similar to that in Ref.~\cite{Aartsen:2015ivb} but with some changes to reduce the energy threshold, as discussed in Ref.~\cite{Aartsen:2014muf,Jakob}.  

First, the number of photoelectrons, $Q_{\rm start}$, required in a rolling 3 $\mu$s time window to define the start of the event, is chosen to depend on the total number of photoelectrons (NPE) of the event, $Q_{\rm tot}$,
 \begin{equation}
  Q_{\rm start}=
  \begin{cases}
    3 & Q_{\mathrm{tot}}<72\,\mathrm{PE}\\
    Q_{\mathrm{tot}}/24 & 72\,\mathrm{PE}\leq Q_{\mathrm{tot}}<6000\,\mathrm{PE}\\
    250 & Q_{\mathrm{tot}}\geq6000\,\mathrm{PE}
  \end{cases}.
\end{equation}

The horizontal veto layer that is placed just below the dust layer is made thicker at 120 m. The veto layer at the bottom is expanded to include the lowest DOM on each string.  This leaves one remaining significant background, from two nearly-coincident air showers.  Sometimes, a low-energy muon can sneak into the detector, producing only a small signal in the veto.  Then, a higher energy muon can enter, producing a high enough number of photoelectrons to satisfy the $Q_{\rm tot}$ requirement.  To avoid this, an algorithm is used to count the number of causally disconnected clusters of light in the detector and reject events where more than  one cluster is found. These cuts reduce the event rate to 0.36 Hz in the 10\% testing sample, compared to an expected starting neutrino rate of 0.20 mHz.  

\begin{figure}
  \centering
  \includegraphics[width=\linewidth]{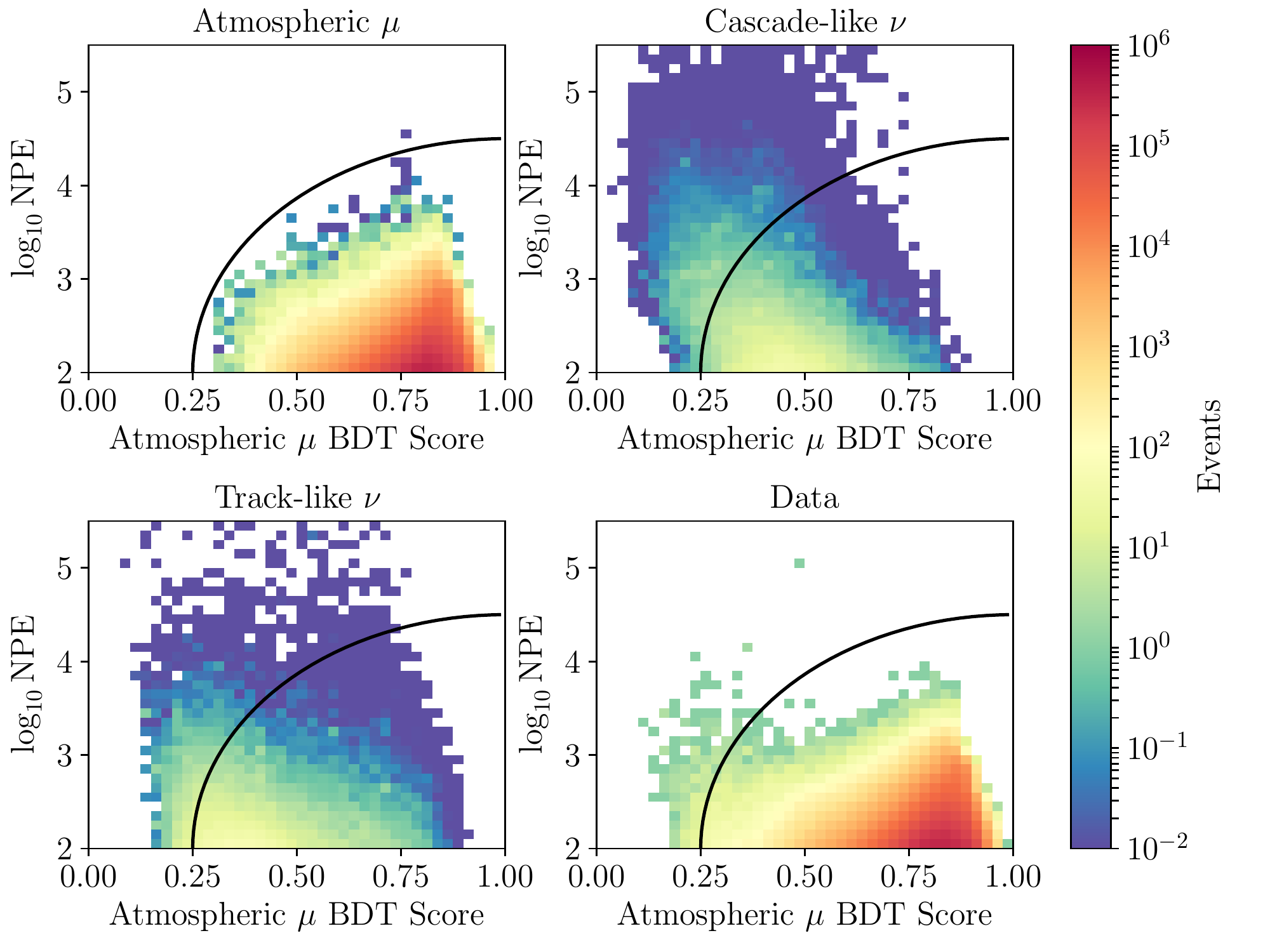}
  \caption{Two-dimensional distributions of atmospheric muon BDT score and number of photoelectrons (NPE), after the outer-layer veto,  for atmospheric muons (top left), cascade-like neutrinos (top right), track-like neutrinos (bottom left) and 10\% of the data (bottom right).    Events above the black lines were accepted as signal events.  To access $\sim 1\,\mathrm{TeV}$ neutrinos that typically produce $\sim 100\,\mathrm{PE}$, one must overcome muon background with a rate 1,800 times higher.}
  \label{fig:bdt}
\end{figure}

\subsection{BDT background rejection}

The remaining background is harder to reject, particularly at lower muon energies where less light is produced.  To further clean up the signal, a boosted decision tree (BDT) is used to classify the passing events as atmospheric muons, starting tracks or cascades.  
The BDT uses 15 variables as input, including $\log_{10}(Q_{\rm tot})$ and the number of photoelectrons in the outer layer veto.  The other variables are from the output of the track and cascade reconstructions.  For the wrong hypothesis ({\it i.e.} tracks reconstructed as cascades), the output may vary greatly, providing considerable separating power.  

The track-reconstruction variables are: cosine zenith of the reconstructed direction, the distance of the track's first visible energy loss from the edge of the detector, the distance between the first and last visible energy loss, the number of direct hits (photoelectrons with DOM arrival time consistent with zero scattering), the track length between the first and last DOMs that registered a direct hit, the estimated angular uncertainty of the track reconstruction, and the angle between the track direction and a simplified reconstruction where first arrival times are fit to a propagating plane wave.  

For the cascade reconstruction, the variables were the depth of the cascade within the detector, the horizontal distance between the cascade and the edge of the detector, the reduced log likelihood of the fit, and the log likelihood ratio of the track and cascade fits.   There are also two variables that use multiple track fits to the same event.  The first considers 104 different in-coming down-going track directions, ending at the reconstructed cascade, and counts the number of photoelectrons in the time window from -15 ns to +1000 ns around the geometric first arrival time from the tracks.  The largest number of photoelectrons among all the tracks is selected for use in the BDT, and this helps to discriminate against down-going muons that do not have a well-reconstructed track.  The second considers 192 outgoing track hypotheses from all directions and counts the number of photoelectrons in the time window from -30 to +500 ns. This is important to identify starting tracks with a low-energy outgoing muon \cite{Jakob,Aartsen:2014muf}. 

The BDT was trained with a sample of 4.5 million simulated atmospheric muon events from CORSIKA \cite{CORSIKA} that pass the outer-layer veto  using SIBYLL2.1  \cite{Ahn2009} to model hadronic interactions.  The spectrum was  weighted to the H3a cosmic-ray flux model \cite{Gaisser:2011cc}.  The neutrino input was a total of
734,000 simulated neutrinos, weighted to the sum of the HKKMS conventional atmospheric flux \cite{Honda:2006qj}, the BERSS prompt neutrino flux \cite{Bhattacharya:2015jpa} and an astrophysical flux with a spectral index $\gamma=2.5$, using the flux from Ref.~\cite{Aartsen:2015zva}.  Several quality criteria were applied to the training sample: neutrinos labeled as starting tracks were required to have a vertex that was contained within the detector and produce a muon having a path length more than 300 m within the detector and energy above 100 GeV.  Further, the reconstructed muon direction was required to be within 5$^o$ of the simulated direction.

\begin{figure}
  \centering
  \includegraphics[width=\linewidth]{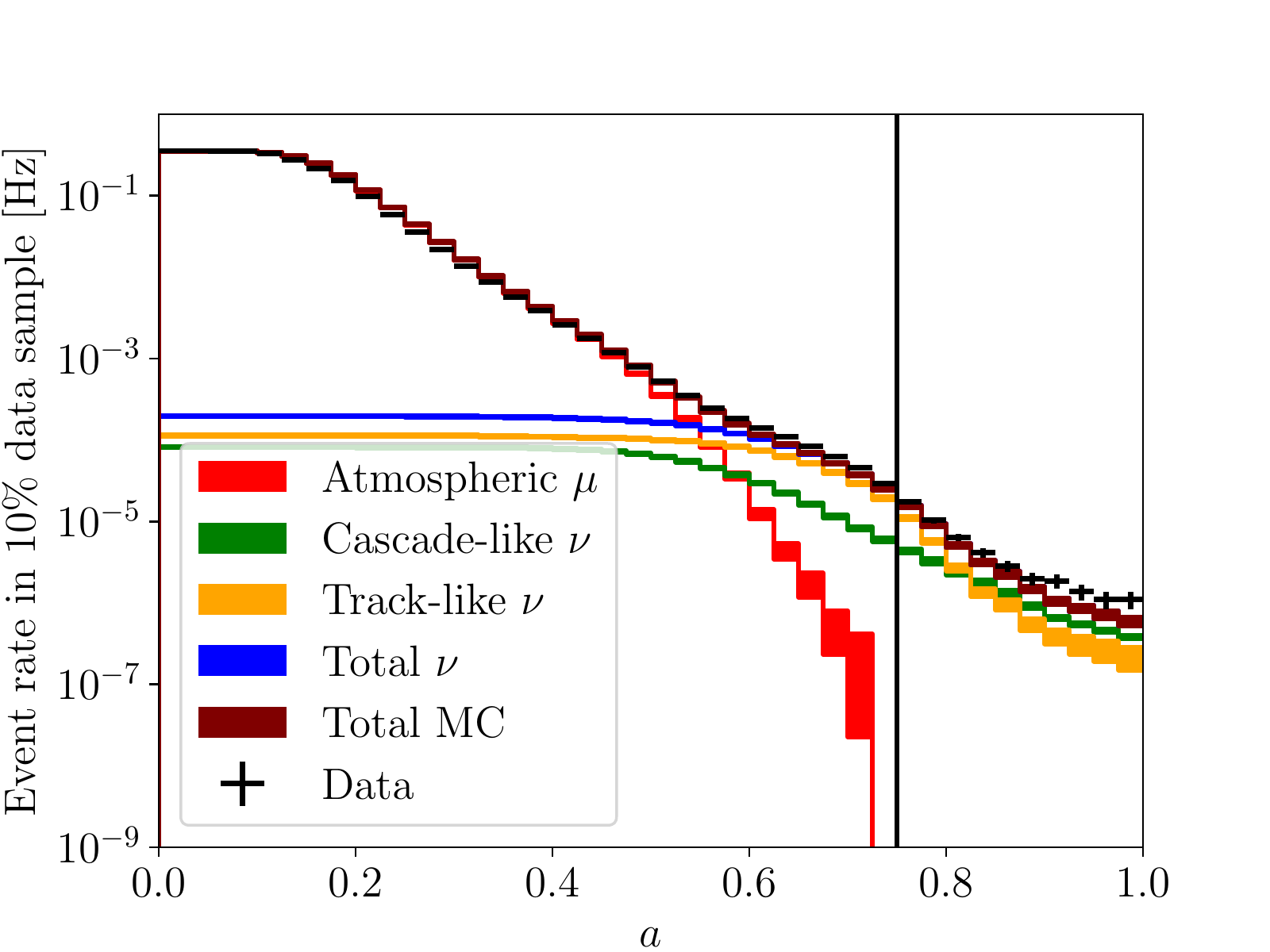}
  \caption{The total event rate for data (black), atmospheric muons (red), cascade-like neutrinos (green) and track-like neutrinos (orange) in the 10\% testing sample as a function of the parameter $a$ appearing in Eq.~\ref{eq:ellipse} describing the elliptical cut.  The chosen value of $a=0.75$ is shown with a vertical black line.}
  \label{fig:ellipse}
\end{figure}

Figure \ref{fig:bdt} shows the number of events as a function of BDT atmospheric muon score and number of photoelectrons,  for atmospheric muons, cascade-like and track-like neutrinos, and data.  There are an estimated 1,800 times as many cosmic-ray muons as neutrinos in the sample, concentrated at low NPE and fairly high BDT muon scores.   To optimally select neutrino events, an elliptical cut was used,
\begin{equation}
  \left(\frac{s_{\mu} - 1}{a}\right)^2 + \left(\frac{\log_{10}(Q_{\rm{tot}}) - 2}{b}\right)^2 > 1,
  \label{eq:ellipse}
\end{equation}  
where $s_{\mu}$ is the BDT atmospheric muon score and $(a,b)$ are parameters describing the semi-major and semi-minor axes of the ellipse.  Values of $a=0.75$ and $b=2.5$ were chosen to eliminate muon background, and the resulting ellipse is shown in Fig.~\ref{fig:bdt}.  The total event rate for data, atmospheric muons, and neutrinos as a function of $a$ in Eq.~\ref{eq:ellipse} while keeping $b$ fixed is shown in Fig.~\ref{fig:ellipse} and shows good agreement between data and simulation.  The chosen value of $a=0.75$ reduces the data event rate to 0.024 mHz or 3615 events in the final data sample.

The CORSIKA Monte Carlo statistics are inadequate to estimate the number of surviving cosmic-ray muons in the final sample; an extrapolation based on the distance from the cut line in Fig. \ref{fig:ellipse} indicates that the number of events should be negligible.  To check this, an additional simulation was made using a parametrization of the flux of single muons in the deep ice calculated from the H3a cosmic-ray model \cite{Jakob}.  The simulation contains 26.8 million single muons passing the outer layer veto, but since it does not contain multi-muon bundles, it underpredicts the total muon background rate at this level by a factor of 3.  However, most events near the BDT cut threshold are single muons, as can be verified by looking at CORSIKA events, so the simulation can still be used to calculate a background estimate.  The predicted background of single muons from the simulation is $2.7\pm 1.0$ events in the final sample.  Even conservatively assuming that an unexpectedly large contribution from bundles increases this by a factor of 3 as observed for the outer layer veto, the resulting muon background of 8.1 events is negligible compared to the full sample size of 3615 events and will no longer be considered.

\begin{figure}
  \centering
  \includegraphics[width=\linewidth]{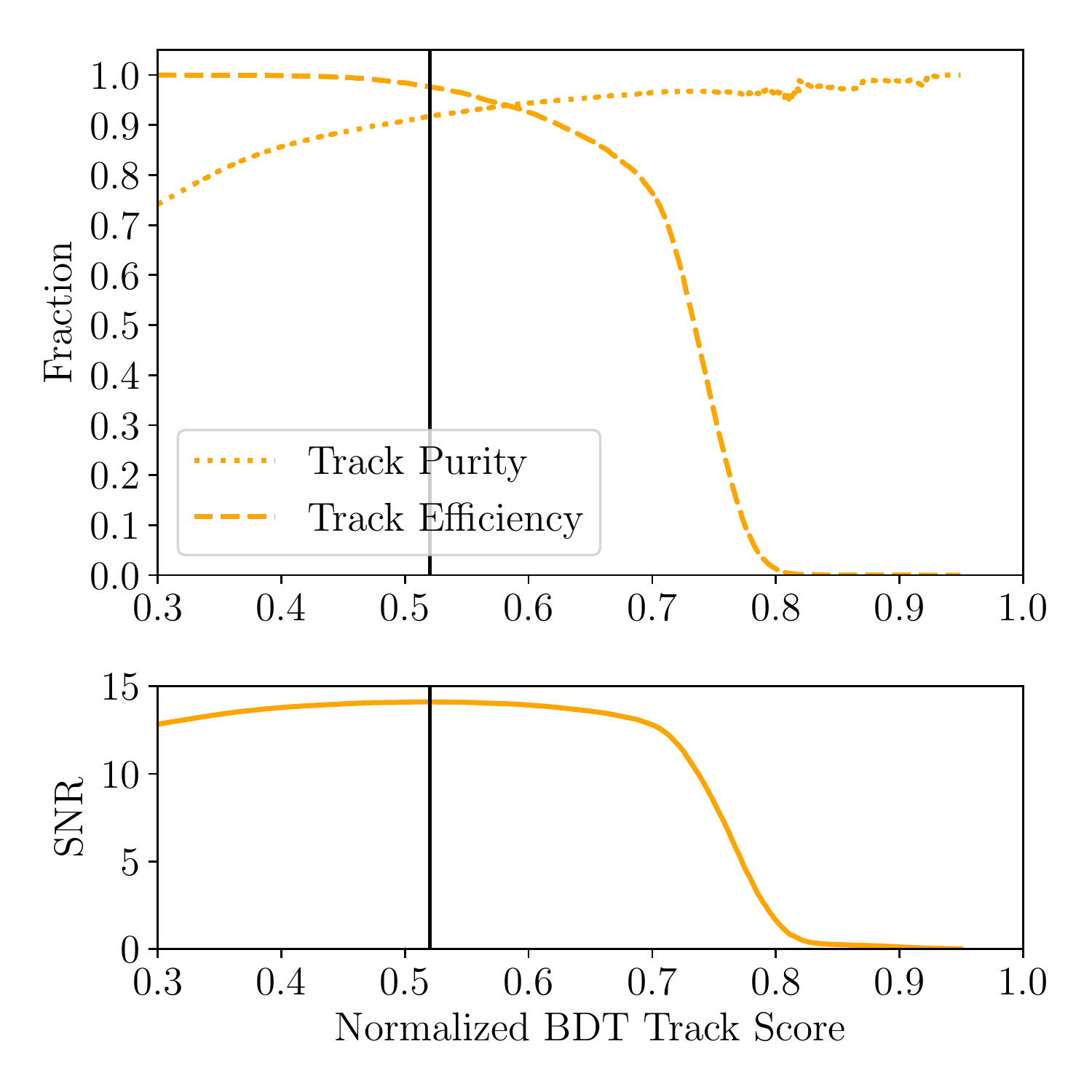}
  \caption{Top: The track purity and efficiency as a function of normalized BDT score $\hat{s}_{\rm{track}}$.  Bottom: The signal-to-noise ratio as a function of $\hat{s}_{\rm{track}}$.
 A BDT track score cut  $\hat{s}_{\rm{track}}  \geq 0.52$ (black line) maximized the signal-to-noise ratio, yielding a 92\% purity and 98\% efficiency.   The fluctuations in the purity curve are due to low statistics in that region.}
  \label{fig:ct_cut}
\end{figure}
 
\subsection{Cascade/track classification}

The same BDT was used to split the neutrino sample into cascades and tracks.  The BDT track score, $s_{\mathrm{track}}$, and cascade score, $s_{\mathrm{casc.}}$ are combined into a variable, $\hat{s}_{\rm{track}} = s_{\mathrm{track}}/(s_{\mathrm{track}}+s_{\mathrm{casc.}})$, which runs from 0 to 1 and is an estimate of an event's ``trackness,'' the likelihood that it contains a track.
The final selection between tracks and cascades depends on the threshold chosen for $\hat s_{\rm track}$.  Figure \ref{fig:ct_cut} shows the purity, efficiency and signal to noise ratio (SNR) as a function of $\hat s_{\rm track}$, as determined using simulated track and cascade samples.  The SNR is defined as the ratio of the true track rate to the size of Poisson fluctuations in the total rate of true tracks and misidentified cascades.  Optimizing SNR, the criterion $\hat{s}_{\rm track} \geq 0.52$ is used to identify tracks.   Events not satisfying this criterion are identified as cascades.

With this classification, there were 965 cascades and 2650 track events in the final sample.   The larger number of tracks is largely due to the dominance of $\nu_\mu$ over $\nu_e$ in the atmospheric neutrino flux. 

\section{Direction and Energy Reconstructions}
\label{sec:reco}

IceCube has previously developed algorithms for reconstructing the direction and energy of cascades  and track events.  Events classified as cascades are reconstructed using a maximum likelihood fit as in Ref.~\cite{Aartsen:2015ivb}, using the full photoelectron timing recorded by each DOM.  Simulations predict that the median angle between the simulated and reconstructed direction is 16$^\circ$.  However, starting tracks have received much less attention and call for new approaches.

\subsection{Starting track reconstruction and inelasticity}

\begin{figure}
  \includegraphics[width=\linewidth]{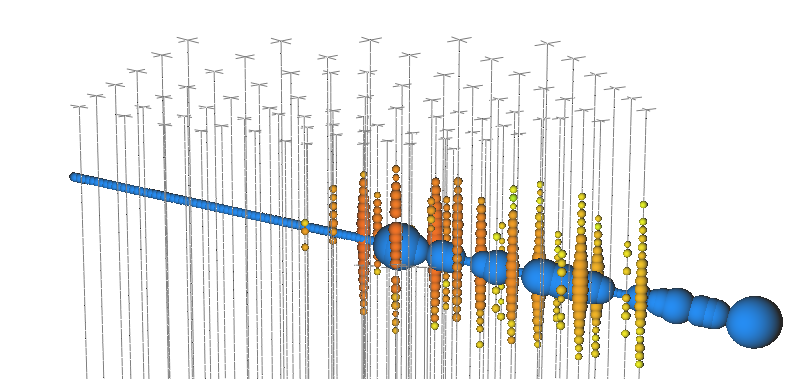}
\vskip .2 in
  \includegraphics[width=\linewidth]{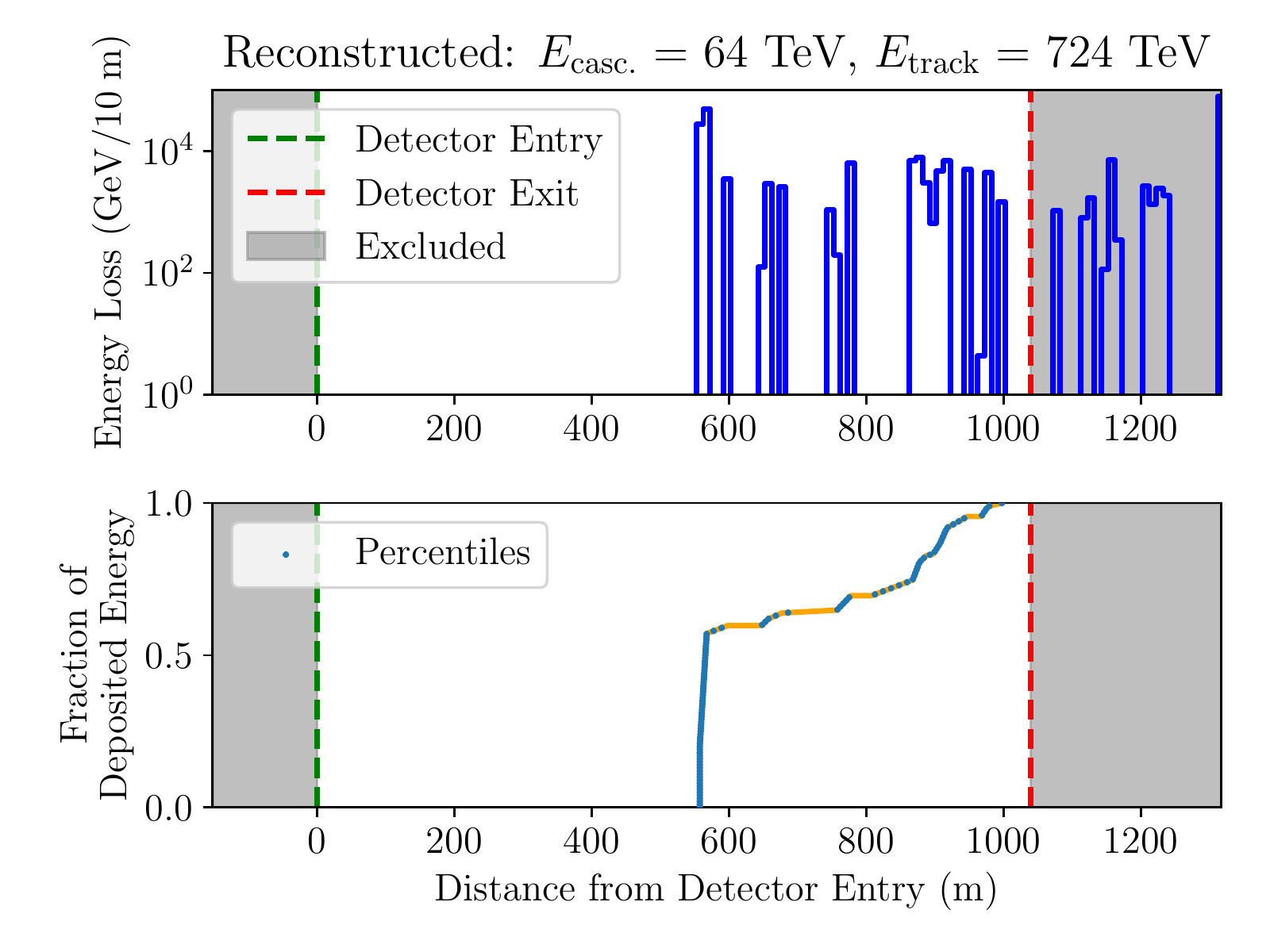}
  \caption{Top: An event display showing the most energetic starting track found in the data.  DOMs are represented by colored spheres with a radius corresponding to the number of photoelectrons detected and with a color showing the first photoelectron time going from red (earliest) to green (latest).  The larger blue spheres show the reconstructed sequence of electromagnetic cascades along the track, and their size is proportional to the reconstructed energy of each cascade on a logarithmic scale.  
 The event originates in the top half of the detector.  Middle: The reconstructed energy loss profile as a function of distance along the reconstructed track.  The detector boundaries are shown by green and red dashed lines.  Energy losses outside the detector, shown in grey, are excluded from the energy and inelasticity reconstruction.  Bottom: The cumulative fraction of the total deposited energy within the detector.  Percentiles of the energy loss distribution, shown with blue points, are features for the random forest regression of cascade and muon energy.  The cascade and muon energies are estimated to be $E_{\mathrm{casc.}}=64\,\mathrm{TeV}$ and $E_{\mu} = 724\,\mathrm{TeV}$, respectively, leading to $E_{\mathrm{vis.}} = 788\,\mathrm{TeV}$ and $y_{\mathrm{vis.}} = 0.08$ for the visible energy and inelasticity, respectively.  The total deposited energy is $135\,\mathrm{TeV}$, and the muon escapes the detector with most of the neutrino's energy.  
}
\label{fig:event} 
\end{figure}

Most IceCube analyses use track reconstructions that fit the track to a straight line by maximizing the likelihood for the reconstruction, based on functions that give the light amplitude and arrival time distribution function for a given DOM, given a track hypothesis \cite{Ahrens:2003ix}.
For starting tracks, these reconstructions have two significant limitations.  First, they assume that the track is infinite, originating outside the detector and traversing entirely through it.  Second, they assume that the muon energy loss is continuous, rather than stochastic.  The latter assumption does not hold for through-going tracks with energies above 1 TeV.  However, it is much more problematic for starting tracks, which are accompanied by a large hadronic cascade from the recoiling nuclear target.  Nevertheless, it is the best reconstruction that we have for finding the directions of starting tracks. 

The directional resolution depends on both the neutrino energy and the inelasticity; the higher the inelasticity, the more the cascade dominates the event.  That said, the median angular error is less than 2$^\circ$ degrees for events with a visible inelasticity up to 0.9, rising to 5$^\circ$  at a visible inelasticity of 0.99. Overall, simulations predict a 1.5$^\circ$  median angular error for the entire starting track sample.  These resolutions do not significantly impact the current analysis.

After the direction is found, the energy loss profile is unfolded as a sequence of electromagnetic cascades along the reconstructed track \cite{Aartsen:2013vja}, as is shown in the middle panel of Fig. \ref{fig:event} for the most energetic starting track found in the sample.  Generally, the largest cascade is at the interaction vertex. 

This unfolding places some of the cascades outside the detector; these cascades have significantly larger uncertainties than those within the detector and are not used for the reconstruction.  The energy loss profile is then integrated to give the cumulative fraction of the energy loss as a function of position along the track, as shown in the bottom panel of Fig.~\ref{fig:event}.   The quantiles where the first 0\%, 1\%, 2\%, {\it etc.}~of total energy loss are determined, where 0\% corresponds to the first observed loss.

\begin{figure}
  \centering
  \includegraphics[width=\linewidth]{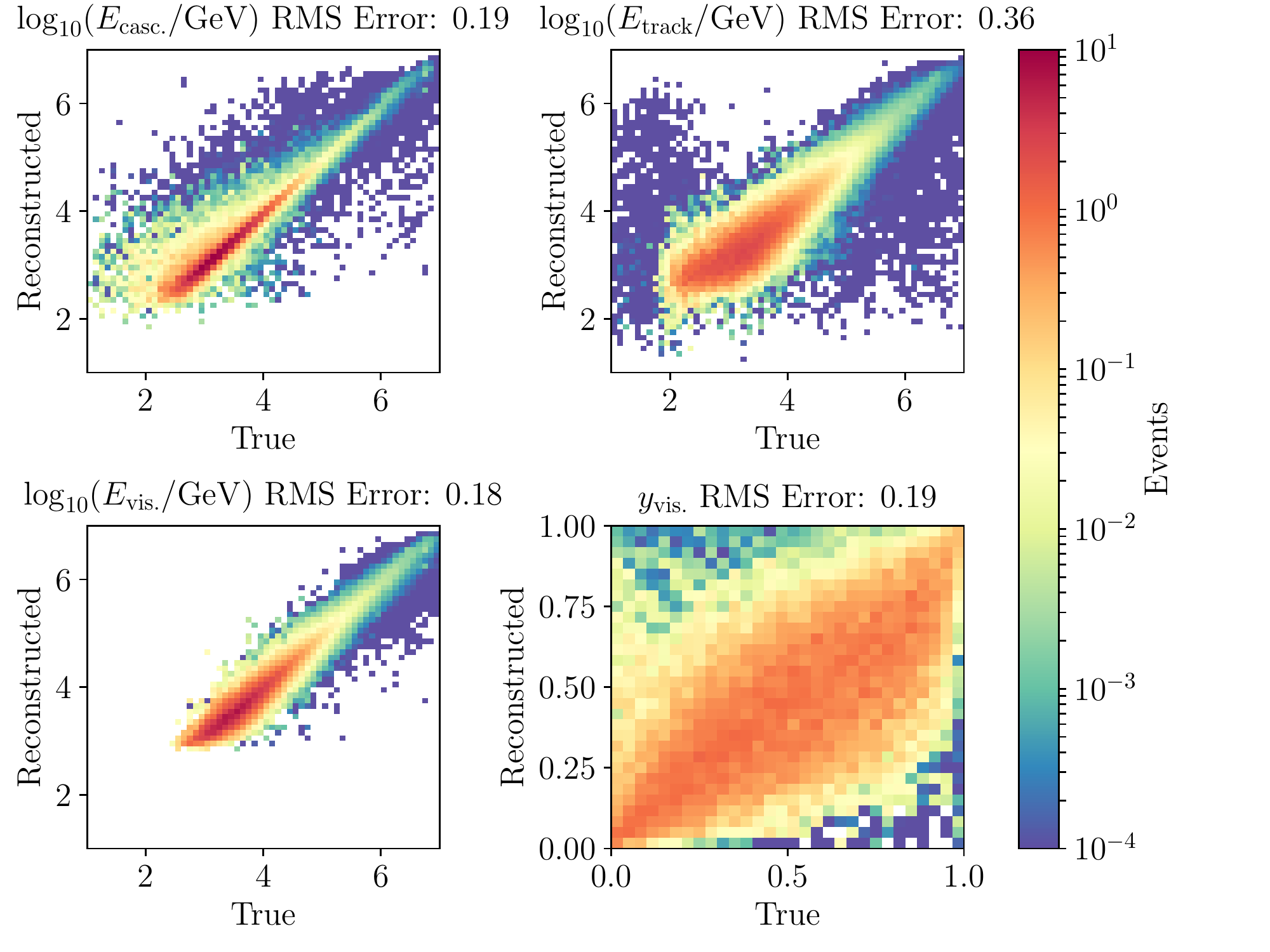}
  \caption{Joint distributions of reconstructed quantity versus true quantity for cascade energy, $E_{\rm casc.}$, muon energy, $E_{\mu}$, total visible energy, $E_{\rm vis.}$, and visible inelasticity, $y_{\rm vis.}$.  The root mean square (RMS) error for each quantity is shown at the top of each panel.}
  \label{fig:ereco_stats}  
\end{figure}

Another machine learning technique, a random forest, is used to estimate the energy of the starting tracks.  It takes the positions of the 101 energy loss quantiles as input, along with the total deposited energy, the total track length contained in the detector and the normalized track BDT score -- a total of 104 inputs.  The random forest is trained using simulated events and validated using an independent sample.   For each event, the forest produces two outputs: the estimated visible cascade energy at the vertex, $E_{\rm casc.}$, and the estimated track energy, $E_{\rm track}$.  These are combined to produce two new variables: the total visible energy, 
\begin{equation}
E_{\rm vis.} = E_{\rm casc.} + E_{\rm track}
\end{equation}
and the visible inelasticity, 
\begin{equation}
y_{\rm vis.} = \frac{E_{\rm casc.}}{E_{\rm vis.}}.
\end{equation}
Since the visible cascade energy is less than the hadronic energy,  $E_{\mathrm{vis.}}$ and $y_{\mathrm{vis.}}$ tend to be lower than the actual neutrino energy and inelasticity, for CC $\nu_{\mu}$ interactions.   Figure \ref{fig:ereco_stats} shows the resolutions for these variables as quantified through the root mean square (RMS) error.  The resolution for $\log_{10} E_{\rm vis.}$ is better than either of its components, because there is some anti-correlation between the cascade and track energies.   With this algorithm, for starting tracks, the RMS error on $\log_{10} E_{\rm vis.}$  is 0.18, better than the typical resolution of 0.22 for through-going muons \cite{Abbasi:2012wht}.  The resolution for the visible inelasticity is 0.19.  These performance metrics are mildly dependent on the assumed neutrino energy spectrum, which here is assumed to follow from the HKKMS conventional atmospheric flux, BERSS prompt atmospheric flux, and the best-fit power-law astrophysical flux from Ref.~\cite{Aartsen:2015zva}.

\begin{figure}
  \centering
  \includegraphics[width=\linewidth]{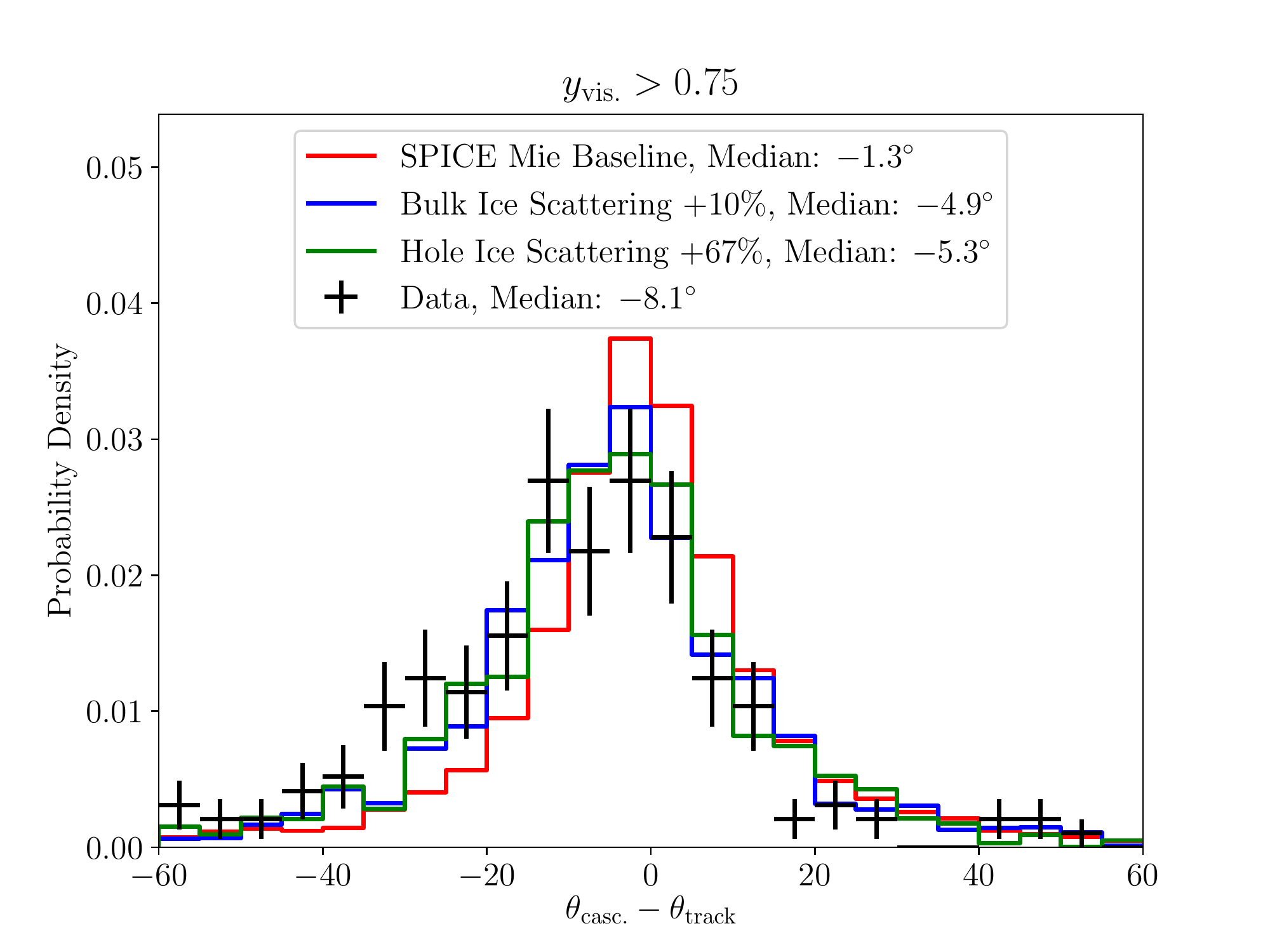}
  \caption{The distribution of the difference in zenith angle between cascade and track reconstructions for tracks with $y_{\mathrm{vis.}}>0.75$.  All reconstructions use the baseline SPICE Mie ice model.  192 events meeting this condition were found in the full 5-year data sample.  The data (black) show a mismatch between cascade and track zenith angles, with cascades being reconstructed with a median angle of 8.1 degrees more downgoing than the corresponding tracks.  When events simulated using the SPICE Mie model are reconstructed (red), the median zenith angle difference is only 1.3 degrees.  This discrepancy may be caused by systematic errors in the ice model since the cascade reconstruction is much more sensitive to the ice model than the track reconstruction.  Distributions for two separate simulations increasing the bulk ice scattering globally by $+10\%$ (blue) and increasing hole ice scattering by $+67\%$ (green) are shown.  Increased bulk ice scattering and hole ice scattering produce a shift that can explain the down-going bias seen in the data.}
\label{fig:ct_ang}  
\end{figure}

\subsection{Cascade angular resolution check}

Starting track events offer an opportunity to study cascade directional reconstruction, using the track as an indicator of true direction.  This is possible because the muon and cascade are boosted to nearly the same direction, and the track angular resolution is much better than the cascade resolution.  Track events with visible inelasticity $y_{\mathrm{vis.}} > 0.75$ are chosen to minimize the effect of the outgoing muon on the cascade reconstruction.  This comparison is sensitive to both systematic offsets, as may be caused by improper modeling of optical scattering in the bulk ice, hole ice, DOMs, or other detector phenomena.  Figure~\ref{fig:ct_ang} shows the distribution of the difference in zenith angle between cascade and track reconstructions for a sample of high-inelasticity tracks.   The data are reconstructed using the SPICE Mie ice model and shows a significant mismatch between the cascade and track zenith angle distributions, with the cascades being reconstructed as more downward-going.  This zenith angle distribution is sensitive to the amount of optical scattering in the bulk ice and in the hole ice, and we find increased scattering can produce a downward bias in the distribution.  Our fits, discussed below, also confirm this observation and find somewhat higher levels of optical scattering than the IceCube baseline.

\section{Inelasticity Fit}

\begin{figure*}
  \centering
  \includegraphics[width=\linewidth]{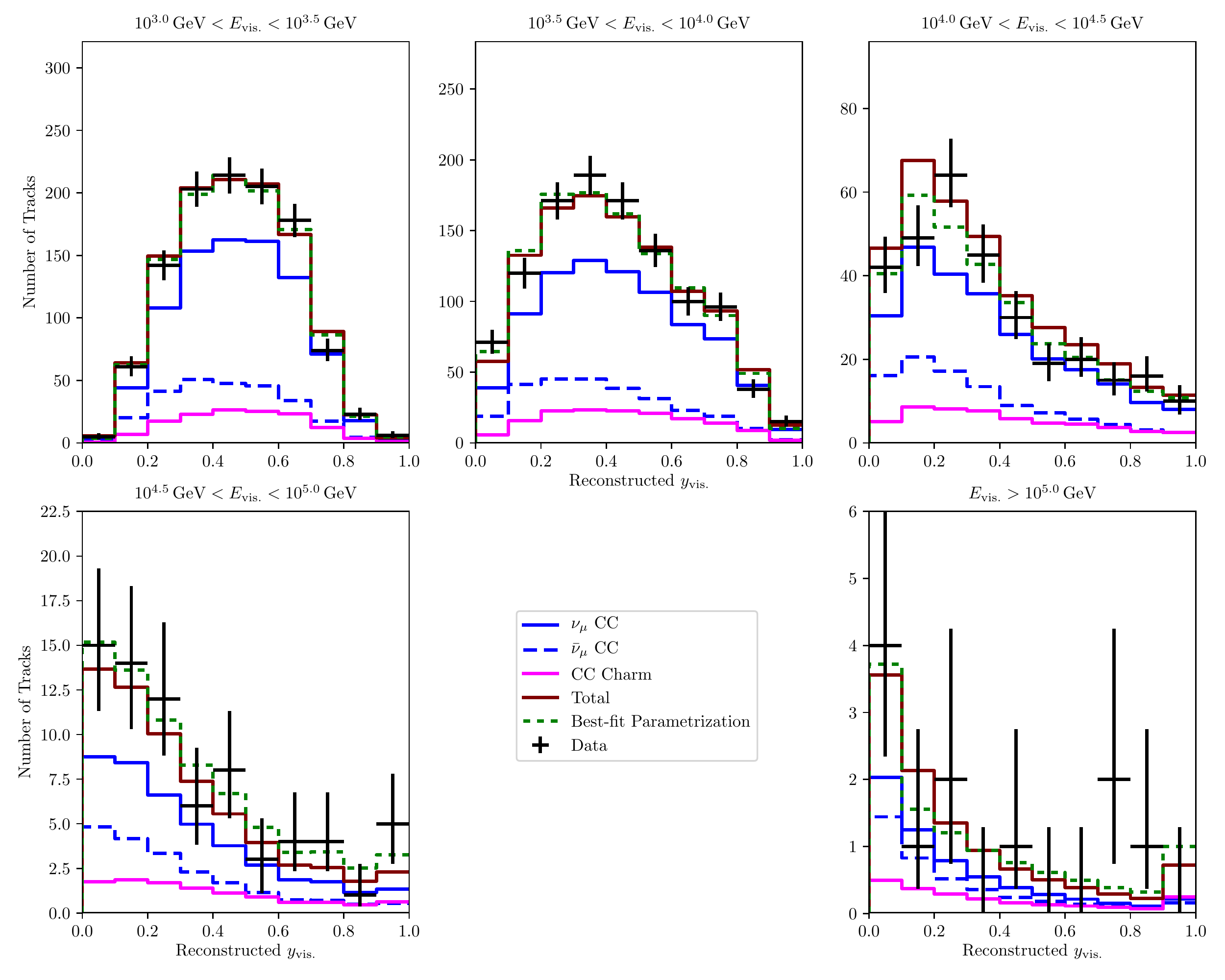}
  \caption{The reconstructed visible inelasticity distribution in five different bins of reconstructed energy.  Observed data are shown in black.  Error bars show 68\% Feldman-Cousins confidence intervals for the event rate in each bin \cite{FC}.  The result of fitting the distribution to the parameterization of Eq.~\ref{eq:nu_cc_xs_y_param} is shown with dashed green lines.  The prediction of the CSMS differential CC cross section are shown for neutrinos with solid blue lines and antineutrinos with dashed blue lines.  The total CC charm contribution is shown in magenta, illustrating its flatter inelasticity distribution.  The best-fit flux models of Tab.~\ref{tab:best_fit_det} are assumed for all predictions.}
  \label{fig:inel_dists}
\end{figure*}

\begin{table}
  \centering
  {\renewcommand{\arraystretch}{1.5}%
  \resizebox{\columnwidth}{!}{%
    \begin{tabular}{|r|r|r|r|r|r|}
    \hline
    $\log_{10}\left(E_{\mathrm{vis.}}/\mathrm{GeV}\right)$ & $\log_{10}\left(E_{\nu}/\mathrm{GeV}\right)$ & Events & $\langle y \rangle$ & $\lambda$ \\
    \hline
    \hline
    $\left[3.0, 3.5\right)$ & $3.33^{+0.20}_{-0.22}$ & 1111 & $0.42^{+0.06}_{-0.09}$ & $1.06^{+0.74}_{-1.90}$ \\
      \hline
      $\left[3.5, 4.0\right)$ & $3.73^{+0.22}_{-0.22}$ & 1107 & $0.42^{+0.02}_{-0.02}$ & $1.09^{+0.25}_{-0.40}$ \\
        \hline
        $\left[4.0, 4.5\right)$ & $4.18^{+0.22}_{-0.22}$ & 310 & $0.38^{+0.03}_{-0.03}$ & $0.97^{+0.26}_{-0.30}$ \\
          \hline
          $\left[4.5, 5.0\right)$ & $4.65^{+0.22}_{-0.22}$ & 72 & $0.37^{+0.05}_{-0.05}$ & $0.75^{+0.44}_{-0.75}$ \\
            \hline
            $>5.0$ & $5.23^{+0.50}_{-0.33}$ & 11 & $0.28^{+0.15}_{-0.24}$ & $0.13^{+0.78}_{-0.13}$ \\
              \hline
  \end{tabular}%
          }
}

\caption{The best-fit parameters when reconstructed inelasticity distributions are fit to Eq.~\ref{eq:nu_cc_xs_y_param} in five bins of reconstructed energy.  The energy range containing the central $68\%$ of simulated neutrino energies for each bin is shown in the second column.  Because of the limited energy resolution, sometimes the 68\% central range extends outside the nominal bin boundaries.}
\label{tab:split_fit_inel}
\end{table}

\begin{figure}
  \centering
  \includegraphics[width=\linewidth]{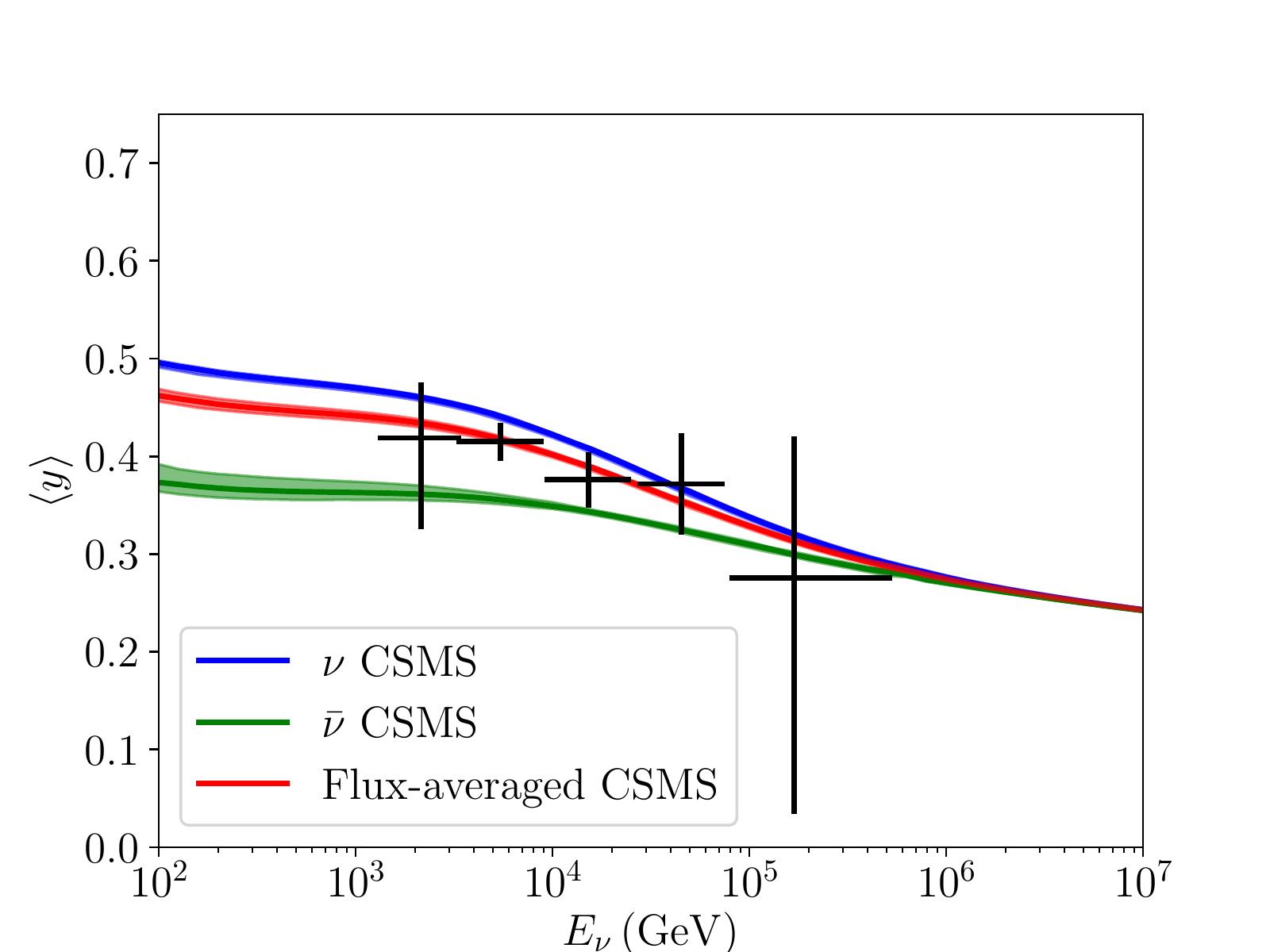}
  \caption{The mean inelasticity obtained from the fit to Eq.~\ref{eq:meany} in five bins of reconstructed energy.  Vertical error bars indicate the $68\%$ confidence interval for the mean inelasticity, and horizontal error bars indicate the expected central $68\%$ of neutrino energies in each bin.  The predicted mean inelasticity from CSMS is shown in blue for neutrinos and in green for antineutrinos.  The height of the colored bands indicates theoretical uncertainties in the CSMS calculation.  A flux-averaged mean inelasticity per the HKKMS calculation is shown in red.}
  \label{fig:split_fit_inel}
\end{figure}

With the reconstruction results from Section \ref{sec:reco}, we can characterize the distribution of visible inelasticity across energy.  The visible inelasticity distribution is shown in Fig.~\ref{fig:inel_dists} for four half-decade energy bins, from 1 TeV to 100 TeV; a 5th bin is used for energies above 100 TeV.  The data are compared to predictions based on the CSMS cross section calculation, weighted by the expected neutrino and antineutrino fluxes.  The flux models used are the best-fit atmospheric and astrophysical models to be described in Sec.~\ref{s:fit}. The data are in good agreement with the predictions.

To further characterize the inelasticity distribution in a model-independent fashion, ideally we would use these visible inelasticity distributions to unfold $d\sigma/dy$, but there are several complications.   The detection efficiency drops at low energies for very large $y$ because the track is no longer visible.  It also drops for small $y$ at low energies because there is not enough light visible in the detector.  Because of these strongly varying efficiencies, the limited statistics and the limited resolution, we do not attempt to unfold the data to present $d\sigma/dy$ distributions.  Instead, we parameterize the true inelasticity distribution, reweight simulation to the parametrized distribution, and fit the visible inelasticity distribution to find the parameters. These parameters can be compared with the Standard Model distribution, and also used to test alternative theories. 

To motivate the parameterization, recall the differential CC cross section can be written schematically at leading order as
\begin{multline}
  \frac{d\sigma}{dxdy}=\frac{2 G_{F}^{2}M E_{\nu} }{\pi}\left(\frac{M_{W}^{2}}{Q^{2}+M_{W}^{2}}\right)^{2} \\ \big[x q(x,Q^2) + (1-y)^2 x \bar{q}(x,Q^2)\big],
  \label{eq:nu_cc_xs_xy_param}
\end{multline}
where $q$ and $\bar{q}$ represent sums of quark PDFs \cite{Connolly:2011vc}.  High-energy neutrinos probe the PDFs at low values of Bjorken$-x\sim3\times 10^{-3}(E_{\nu}/\mathrm{PeV})$, where sea quarks dominate, and they should have a power-law behavior, $xq(x,Q^2)\sim A(Q^2)x^{-\lambda}$ with $\lambda \sim 0.4$.  Following Ref.~\cite{Pena2001}, when transforming variables from $(x,y)$ to $(Q^2,y)$, the $Q^2$-dependence of Eq.~\ref{eq:nu_cc_xs_xy_param} can be separated from the $y$-dependence and integrated out to give a two-parameter function, 
\begin{equation}
  \frac{d\sigma}{dy} \propto \left(1 + \epsilon (1-y)^2\right)y^{\lambda - 1},
  \label{eq:nu_cc_xs_y_param}
\end{equation}
where the parameter $\epsilon$ indicates relative importance of the term proportional to $(1-y)^2$ in Eq.~\ref{eq:nu_cc_xs_y_param}.  This parameterization also works for antineutrinos, but $\epsilon$ takes on a different value since $q$ and $\bar{q}$ are interchanged.  Our measurement represents an average over neutrinos and antineutrinos.  The normalized inelasticity distribution can then be written as
\begin{equation}
  \frac{dp}{dy} = N \left(1 + \epsilon (1-y)^2\right)y^{\lambda - 1},
  \label{eq:nu_cc_dpdy_param}
\end{equation}
where $N$ is the normalization
\begin{equation}
  N = \frac{\lambda(\lambda+1)(\lambda+2)}{2\epsilon + (\lambda+1)(\lambda+2)}.
\end{equation}
This simple parameterization can accurately represent sophisticated calculations of inelasticity distributions.  For example, a fit of Eq.~\ref{eq:nu_cc_xs_y_param} to the full CSMS calculation produces no more than a $1\%$ root mean square deviation (averaged over $y$) for neutrino energies from $1\,\mathrm{TeV}$ to $10\,\mathrm{PeV}$.

In practice, the parameters $\epsilon$ and $\lambda$ are highly correlated when fitting Eq.~\ref{eq:nu_cc_dpdy_param} to realistic inelasticity distributions.  To avoid this correlation, it is convenient to fit for the mean of the distribution, $\langle y \rangle$, and $\lambda$ instead, which show far less correlation.  The mean inelasticity can be found analytically,
\begin{equation}
  \langle y \rangle = \int_0^1 y \frac{dp}{dy}dy = \frac{\lambda (2\epsilon + (\lambda + 2)(\lambda + 3))}{(\lambda + 3)(2\epsilon + (\lambda + 1)(\lambda + 2))}
  \label{eq:meany}
\end{equation}
It is then straightforward to substitute 
\begin{equation}
  \epsilon = -\frac{(\lambda+2)(\lambda+3)}{2}\frac{\langle y \rangle(\lambda + 1) - \lambda}{\langle y \rangle(\lambda + 3) - \lambda}
\end{equation}
into Eq.~\ref{eq:nu_cc_dpdy_param} so that $dp/dy$ can be found as a function of $\langle y\rangle$ and $\lambda$ only.

The visible inelasticity distribution in each energy range from Fig.~\ref{fig:inel_dists} can then be fit to the parametrization of Eq.~\ref{eq:nu_cc_dpdy_param} using a binned Poisson likelihood fit of the 10 bins.  The goodness-of-fit test statistic is
\begin{equation}
  -2\ln \Lambda = 2 \sum_i \left[\mu_i(\boldsymbol{\theta}) - n_i + n_i \ln\frac{n_i}{\mu_i(\boldsymbol{\theta)}}\right] + \sum_j \frac{(\theta_j - \theta^{*}_j)^2}{\sigma_j^2}
  \label{eq:llh}
\end{equation}
where $\mu_i(\boldsymbol{\theta})$ is the expected event count in each bin depending on parameters $\boldsymbol{\theta}$ and $n_i$ is the observed event count per bin \cite{PDG}.  The second sum accounts for a Gaussian prior distribution on a parameter $\theta_j$ with mean $\theta^{*}_j$ and standard deviation $\sigma_j$.  The expected event rate is derived from weighted Monte Carlo simulations.  To account for the parametrized inelasticity distribution from Eq.~\ref{eq:nu_cc_dpdy_param}, each simulated event receives a reweighting factor, $\frac{dp}{dy}(y;\langle y \rangle,\lambda)/\frac{dp}{dy}_{\rm{CSMS}}$, where $\frac{dp}{dy}_{\rm{CSMS}}$ is the inelasticity distribution calculated by CSMS that is used in the simulation.  A total event rate scaling factor is also included to account for uncertainties in the flux normalization.  The neutrino flux is assumed to follow the best-fit flux models in Sec.~\ref{s:fit}, but the flux model and its uncertainties have negligible effect since the size of each energy range is comparable to the energy resolution.  Detector systematic uncertainties on bulk ice scattering and absorption, DOM optical efficiency, and hole ice scattering are included through the use of 4 additional nuisance parameters in the fit.  They are constrained by Gaussian priors and are further described in Sec.~\ref{s:fit}.

The best-fit parameters for describing these inelasticity distributions are shown in Tab.~\ref{tab:split_fit_inel}.  Figure \ref{fig:split_fit_inel} compares $\langle y \rangle$ as a function of energy with the predictions of the CSMS calculation for neutrinos and antineutrinos.  The measured values of $\langle y \rangle$ agree well with the flux-weighted average of neutrinos and antineutrinos.  The downward trend in $\langle y \rangle$ is due to the $W$-boson propagator.

\section{Likelihood fit results and starting track/cascade comparison}
\label{s:fit}

\begin{figure*}
  \centering
  \includegraphics[width=\linewidth]{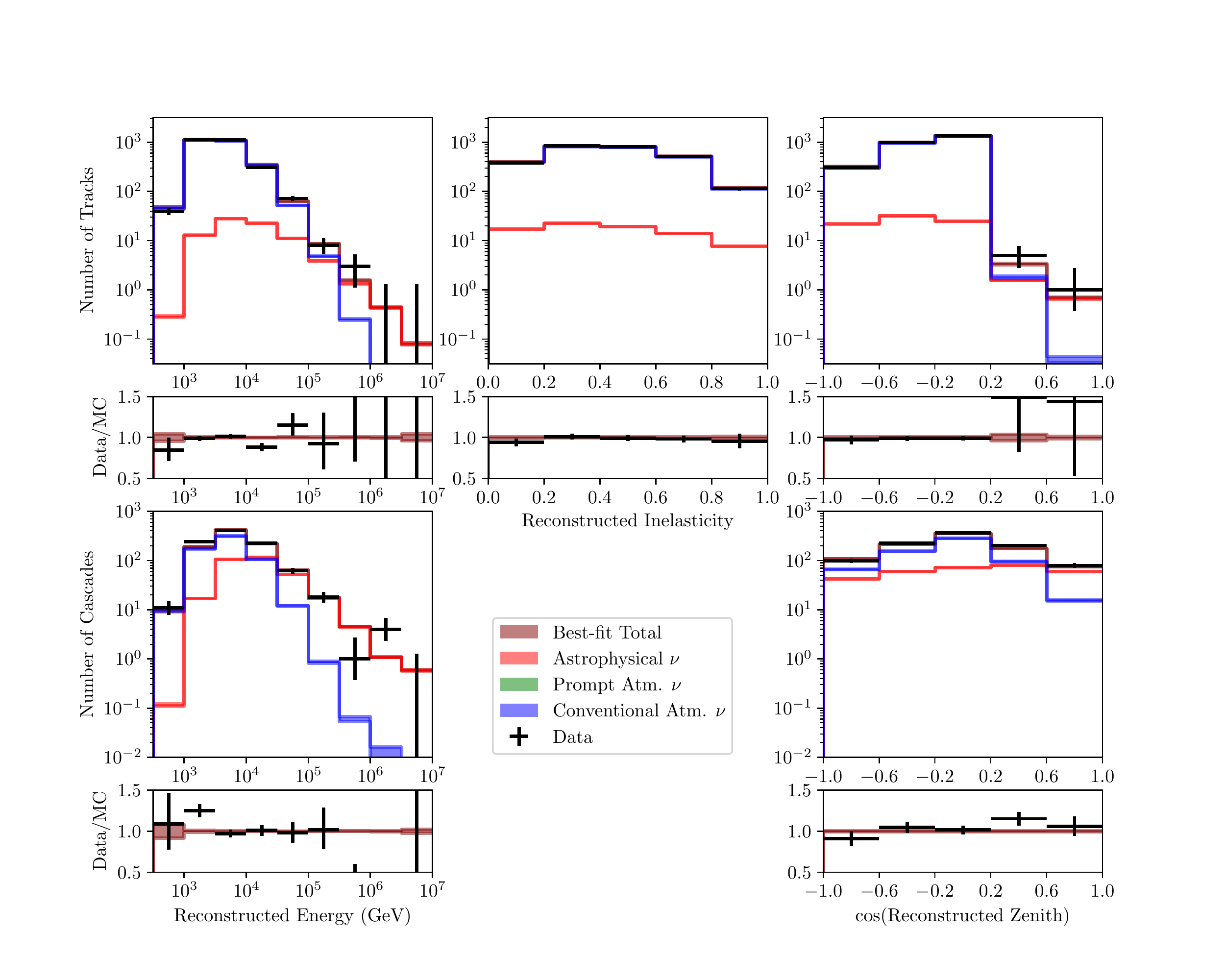}
  \caption{Best-fit distributions with all neutrino flux and detector parameters.  Top: The distribution of reconstructed visible energy, visible inelasticity, and cosine zenith for the sample of starting tracks.  Bottom: the distribution of reconstructed energy and cosine zenith for the sample of cascades.  Contributions of conventional atmospheric and astrophysical neutrinos are shown in blue and red, respectively, and the total predicted distribution is shown in maroon.  The prompt atmospheric neutrino contribution is not shown since its best-fit value is zero.  The black error bars show the data.  The inclusion of detector parameters describing bulk and hole ice scattering substantially improves the model description of the cascade cosine zenith distribution.  The best-fit parameters are shown in Tab.~\ref{tab:best_fit_det}.}
  \label{fig:3d_dists_det}  
\end{figure*}

\begin{table}
  \centering
  {\renewcommand{\arraystretch}{1.5}%

  \resizebox{\columnwidth}{!}{%
\begin{tabular}{|r|r|r|}                                                       
\hline                                                                         
Parameter & $68\%$ CI & Prior \\                                               
\hline                                                                         
\hline                                                                         
$\Phi_{\mathrm{conv}}$ & $1.05^{+0.07}_{-0.06}$ & Flat $[0,\infty)$ \\                      
\hline                                                                         
$\Delta\gamma_{\mathrm{cr}}$ & $0.04^{+0.03}_{-0.03}$ & Gaussian $0.00 \pm 0.05$ \\
\hline
$R_{K/\pi}$ & $1.11^{+0.35}_{-0.28}$ & Gaussian $1.00 \pm 0.50$ \\
\hline
$\Phi_{\mathrm{prompt}}$ & $0.00^{+1.10}_{-0.00}$ & Flat $[0,\infty)$ \\
\hline
$\Phi_{0}/10^{-18}\,\mathrm{GeV}^{-1}\mathrm{s}^{-1}\mathrm{cm}^{-2}\mathrm{sr}^{-1}$ & $2.04^{+0.23}_{-0.21}$ & Flat $[0,\infty)$ \\
\hline
$\gamma$ & $2.62^{+0.07}_{-0.07}$ & Flat \\
\hline
$\epsilon_{\rm DOM}/\epsilon_{0}$ & $1.08^{+0.03}_{-0.03}$ & Gaussian $1.00 \pm 0.10$ \\
\hline
$\alpha_{\mathrm{Scat.}}$ & $1.07^{+0.01}_{-0.01}$ & Gaussian $1.00 \pm 0.10$ \\
\hline
$\alpha_{\mathrm{Abs.}}$ & $1.02^{+0.02}_{-0.02}$ & Gaussian $1.00 \pm 0.10$ \\
\hline
$\alpha_{\mathrm{Hole\;Ice}}$ & $1.46^{+0.12}_{-0.12}$ & Flat $[1.00,1.67]$ \\
\hline
\hline
Test Statistic $-2 \ln \Lambda $ &  175.54 & -- \\
\hline
\end{tabular}%
}
}
\caption{The best-fit parameters including all neutrino flux and detector systematic uncertainties.  $68\%$ confidence intervals are shown as calculated by the profile likelihood method.  The prior distribution for each parameter is shown in the last column.  The last row is the goodness-of-fit test statistic.}
\label{tab:best_fit_det}  
\end{table}

\begin{figure}
  \centering
  \includegraphics[width=\linewidth]{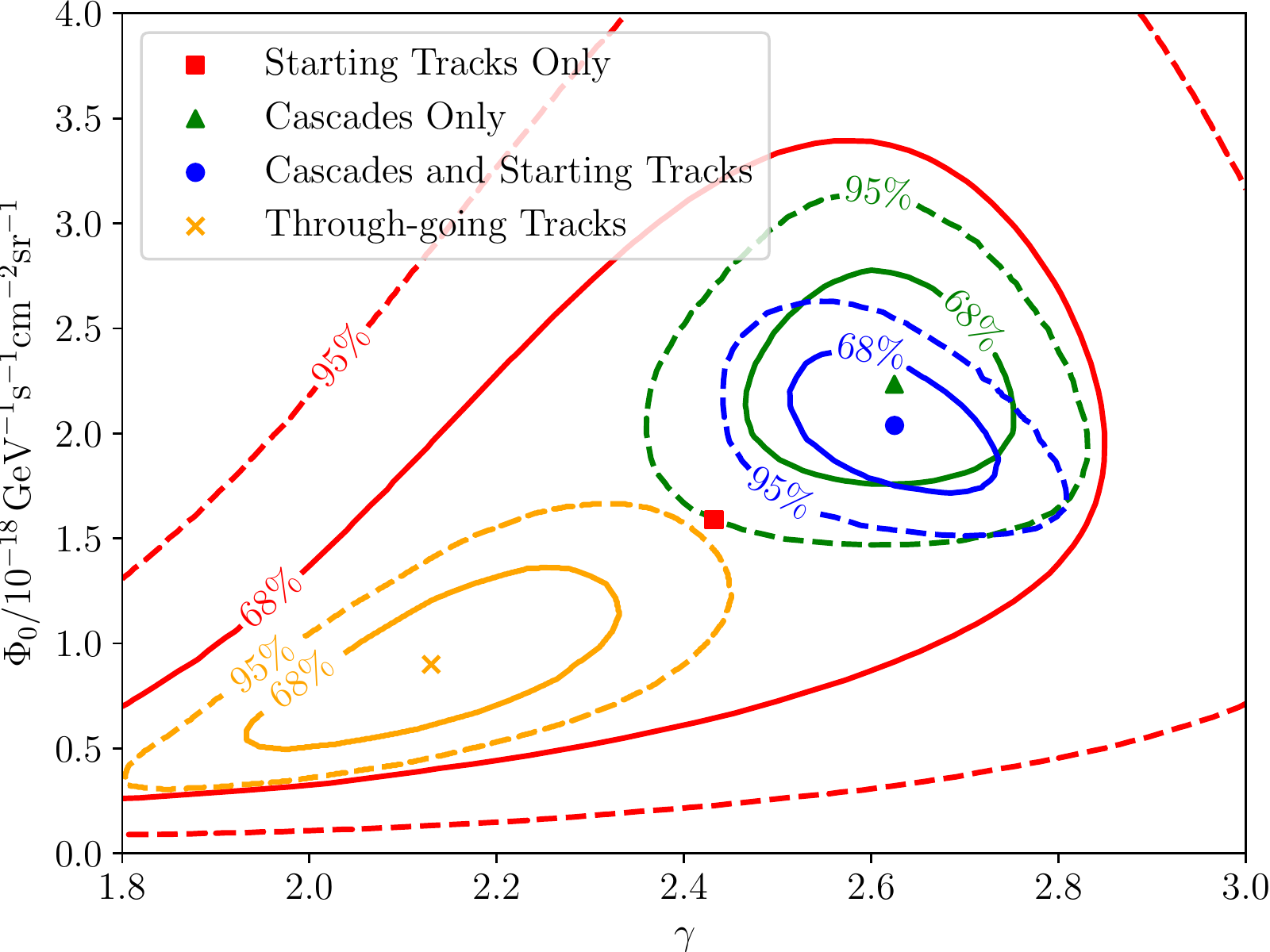}
  \caption{Confidence regions for the astrophysical power-law index, $\gamma$ and flux normalization, $\Phi_0$.  The blue contours show the confidence region for the joint fit of the cascade and starting track samples, which is the main result obtained here.  The red contours show the confidence region for a fit of starting tracks only, and the green contours show the confidence region for a fit of cascades only, and all are consistent with each other.  The confidence region from the IceCube analysis of through-going tracks \cite{Aartsen:2016xlq} is shown in orange, which is in tension with the cascade confidence region.  The contours from the starting track sample are consistent with both cascade and through-going samples.}
  \label{fig:astro_2d_scan}
\end{figure}

Inelasticity introduces a new dimension into the studies of high-energy astrophysical neutrinos, providing sensitivity to a number of new phenomena.   In addition, the much more precise measurement of the starting tracks allows new tests, such as comparing the energy spectra of astrophysical $\nu_\mu$ with that from cascades, a mixture that is mostly $\nu_e$ and $\nu_{\tau}$.  The inelasticity distribution is also sensitive to the $\nu/\overline\nu$ ratio and to neutrino interactions that produce charm quarks.  In this section, we will present a baseline maximum likelihood fit and compare it with previous analyses.

The fit is done jointly over both cascades and starting tracks.  For cascades, data is binned in two dimensions with half decade bins in energy ranging from $10^{2.5}\;\rm{GeV}$ to $10^{7}\;\rm{GeV}$ and 5 bins in cosine zenith angle.  For tracks, data is binned in three dimensions with the same energy and zenith binning but additionally 5 bins in visible inelasticity.  The same binned Poisson test statistic in Eq.~\ref{eq:llh} is used.

We first fit the data to a model that is similar to previous IceCube analyses \cite{Aartsen:2013jdh,Aartsen:2014gkd,Aartsen:2015zva,Aartsen:2015ivb}.  It includes three components: the conventional atmospheric neutrino flux, the prompt atmospheric flux and the astrophysical flux.  After describing this standard fit here, Sections VIII.A-VIII.D discuss some additional fits which each add one additional degree of freedom to explore additional aspects of the physics. 

The conventional atmospheric flux is based on the HKKMS calculation, extrapolated in energy to above 10 TeV and modified to include the knee of the cosmic-ray spectrum following the H3a cosmic-ray model.  To account for the uncertainties in this flux model, we include several nuisance parameters in the fit.  The first is the overall normalization, $\Phi_{\rm conv}$.  The second, $\Delta\gamma_{\rm cr}$, accounts for uncertainty in the energy spectrum by allowing the spectral index to vary with a prior.  A third parameter, $R_{K/\pi}$, accounts for uncertainties in the kaon to pion ratio in cosmic-ray air showers \cite{Aartsen:2015xup}.   Neutrinos from kaons have a somewhat different zenith angle distribution than those from pions; $R_{K/\pi}$ accounts for that possible variation.    The prompt atmospheric flux follows the BERSS calculation, an update of the ERS calculation \cite{Enberg:2008te} used in many previous IceCube works.  It is incorporated into the analysis with a single parameter, the normalization for the overall amplitude.  The self-veto probability is included for both atmospheric flux calculations.

Astrophysical neutrinos are initially assumed to be isotropic.  In this section the $\nu_e:\nu_\mu:\nu_\tau$ ratio is taken to be $\left(\frac{1}{3}:\frac{1}{3}:\frac{1}{3}\right)_{\oplus}$, an approximation expected from almost any conventional source model, after accounting for oscillation en-route to Earth.  The $\nu:\overline\nu$ ratio is taken to be $1:1$. The flux per flavor is assumed to follow a single power-law:
\begin{equation}
\Phi_\alpha(E_{\nu}) = 3 \Phi_0 f_{\alpha,\oplus} \left(\frac{E_{\nu}}{\rm 100\,TeV}\right)^{-\gamma}
\end{equation}
where $f_{\alpha,\oplus} \approx 1/3$ is the fraction of each flavor at Earth, $\gamma$ is the power-law index, and $\Phi_0$ is a normalization factor that corresponds to the average flux of $\nu$ and $\bar{\nu}$ per flavor at 100 TeV.

Detector systematic uncertainties are incorporated in all results through the use of four more nuisance parameters that describe uncertainties in the detection and transmission of light through ice.  The first, $\epsilon_{\rm DOM}$, accounts for uncertainties in the overall optical sensitivity of the DOMs; the prior on this is $\pm 10\%$.  Two parameters, $\alpha_{\rm Scat.}$ and $\alpha_{\rm Abs.}$, account for uncertainties of optical scattering and absorption in the bulk ice.  These parameters linearly scale the inverse scattering and absorption lengths uniformly over all ice layers.   Finally, $\alpha_{\rm Hole\;Ice}$ accounts for uncertainties on the overall scattering in the hole ice, the columns of refrozen drill water that the strings are emplaced in.  Because of the presence of visible air bubbles \cite{Rongen:2016sbk} and possible impurities, the optical quality of this ice is expected to be much worse than that of the rest of the ice.   The baseline ice model assumes a 50 cm scattering length in hole ice, but calibration data are also consistent with a scattering length of 30 cm, or 1.67 times more scattering.  Uncertainties in both the hole ice and bulk ice scattering can lead to a bias in the zenith angle distribution of cascades as shown in Fig.~\ref{fig:ct_ang}.   The fit finds values for these parameters that are in line with expectations. The larger value of $\alpha_{\rm Hole\;Ice}$ is comparable with other recent IceCube measurements \cite{Aartsen:2017mau}. 

The fit results are shown in Tab.~\ref{tab:best_fit_det}, and the fit is compared with the data in Fig. \ref{fig:3d_dists_det}.  Based on a set of computed pseudo-experiments, the goodness-of-fit test statistic, $-2\ln{\Lambda}=175.54$, corresponds to a probability ($p$-value) of 0.04.  This value indicates the fit model may not be a complete description of the data, perhaps due to inadequately simulated systematic uncertatinties.  Still, the $p$-value is acceptable, and there are no obvious problem areas visible in the distributions in Fig. \ref{fig:3d_dists_det}.  We also fit the data with the optical properties of the ice fixed to their default values.  The test statistic worsened significantly, but, except for $R_{K/\pi}$, the other fit parameters did not change significantly.  The inelasticity measurements are quite robust against systematic errors.

\subsection{Astrophysical neutrino energy spectrum}

The baseline fit finds an astrophysical power law index of $\gamma=2.62\pm 0.07$, with a normalization $\Phi_0 = 2.04^{+0.23}_{-0.21} \times 10^{-18}\,\mathrm{GeV}^{-1}\mathrm{s}^{-1}\mathrm{cm}^{-2}\mathrm{sr}^{-1}$.  This is in agreement with earlier IceCube studies of contained events \cite{Aartsen:2014gkd}, but softer than the most recent measurements of contained events \cite{Aartsen:2017mau}, and
considerably steeper than the measurement using through-going tracks, which found $\gamma=2.13\pm 0.13$ \cite{Aartsen:2016xlq}. There are a couple of possible explanations for the latter tension.  First, the through-going tracks have considerably higher neutrino energies than the current sample, with 90\% of the sample sensitivity in the energy range from 194 TeV to 7.8 PeV.  Using the same method as in Ref.~\cite{Aartsen:2016xlq}, the current sample has a 90\% central range of 3.5 TeV to 2.6 PeV.  For reference, 90\% of the selected contained events are in the energy range from 3.3 TeV to 220 TeV; this range is much lower than for the sensitivity because the most energetic events have the largest effect on the astrophysical flux measurement.   If the astrophysical flux is not a single power law, then one might measure different spectral indices in different energy ranges.  Or, with difficulty, one might find different spectral indices for tracks and cascades. 
One way to test this hypothesis is to repeat the fit, allowing the astrophysical spectral indices and normalizations to vary between the starting tracks and cascades.  When this is done, we find a power-law index of $\gamma=2.43^{+0.28}_{-0.30}$ for starting tracks and $\gamma=2.62\pm 0.08$ for the cascades.  The two indices are compatible within uncertainties.  Figure \ref{fig:astro_2d_scan} shows the two-dimensional confidence regions for the cascade and track measurements, the combined measurement, and the previous through-going track fit. Confidence regions are derived from the profile likelihood over all nuisance parameters, and it is assumed the test statistic follows a $\chi^2$ distribution throughout.  The cascade sample drives the combined-sample index of $2.62\pm0.07$, by virtue of the much better energy resolution and lower atmospheric background compared to tracks.  Within the uncertainty, the starting track power-law index is also compatible with that from the through-going tracks.  We considered alternate scenarios with a double power-law or a power law with a cutoff that could explain a harder power-law index found at high energies, but we found no evidence for either when fitting our sample alone.  All later results will continue the assumption of an unbroken power-law spectrum. 

The other parameters in the fit in Tab.~\ref{tab:best_fit_det} are in line with expectations.  The best-fit prompt flux is zero, in agreement with many previous IceCube studies \cite{Aartsen:2014gkd,Aartsen:2014muf,Aartsen:2015knd,Aartsen:2016xlq}, but the $1\sigma$ upper limit is compatible with the expected BERSS flux.  The conventional flux, cosmic-ray spectral index and $R_{K/\pi}$ are all in line with expectations.

For the remainder of this section, we will add parameters one at a time to this baseline fit to independently measure the flavor composition of astrophysical neutrinos, the atmospheric $\nu_{\mu}:\bar{\nu}_{\mu}$ ratio, and neutrino charm production.  The results are little affected if we choose allow all of these parameters float simultaneously, and none of these measured parameters show strong correlation with another.

\subsection{Astrophysical neutrino flavor composition}

\begin{figure}
  \centering
  \includegraphics[width=1.1\linewidth]{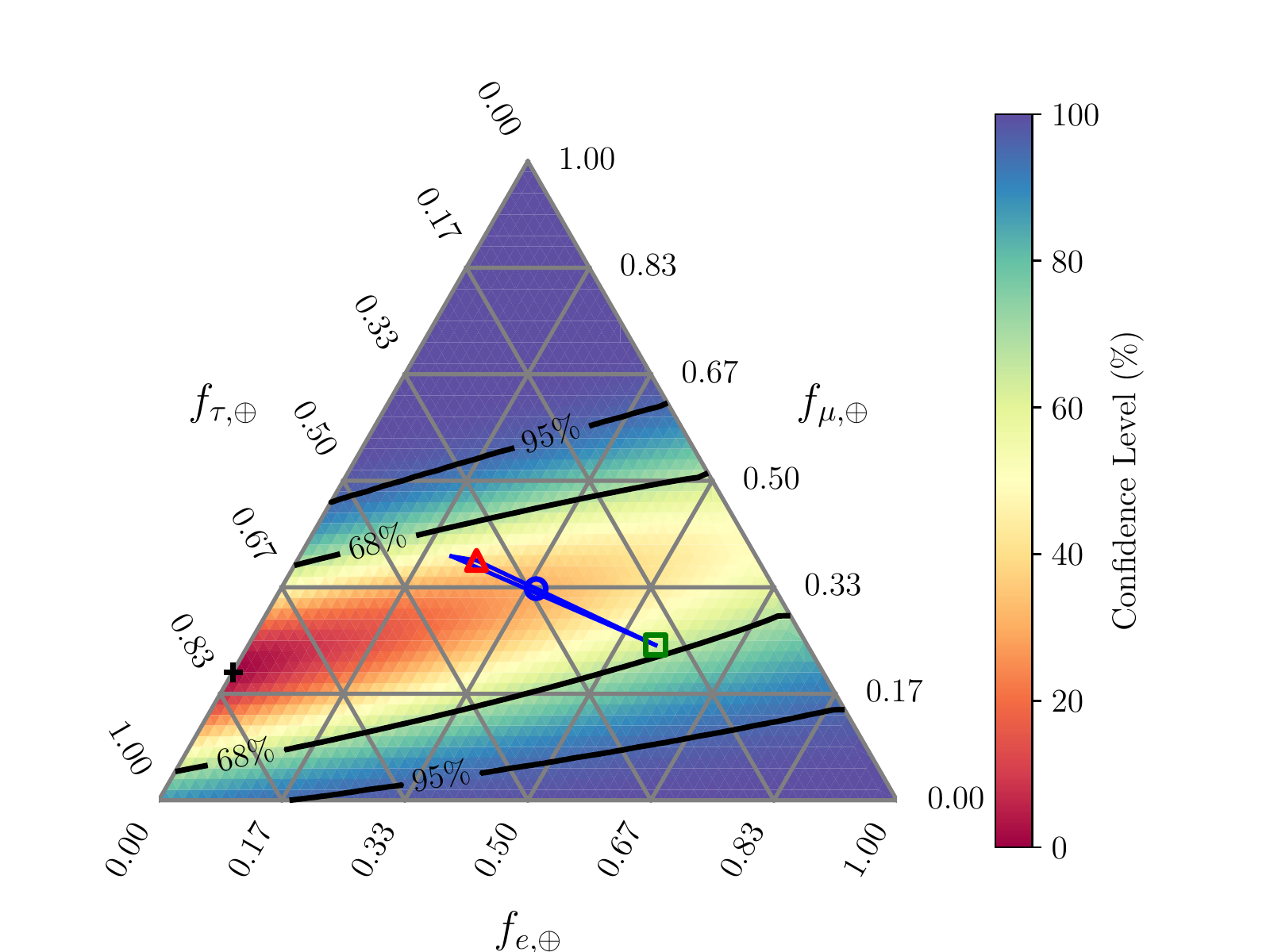}
  \caption{Confidence regions for astrophysical flavor ratios $(f_{e} \mathbin{:} f_{\mu} \mathbin{:} f_{\tau})_{\oplus}$ at Earth. The labels for each flavor refer to the correspondingly tilted lines of the triangle. Averaged neutrino oscillations map the flavor ratio at sources to points within the extremely narrow blue triangle diagonally across the center. The $\approx \left(\frac{1}{3}\mathbin{:} \frac{1}{3} \mathbin{:} \frac{1}{3}\right)_{\oplus}$ composition at Earth, resulting from a $\left(\frac{1}{3} \mathbin{:} \frac{2}{3} \mathbin{:} 0\right)_S$ source composition, is marked with a blue circle. The compositions at Earth resulting from source compositions of $\left(0 \mathbin{:} 1 \mathbin{:} 0\right)_S$ and $\left(1 \mathbin{:} 0 \mathbin{:} 0\right)_S$ are marked with a red triangle and a green square, respectively.  The updated best-fit neutrino oscillation parameters from \cite{Esteban2017} are used here.  Though the best-fit composition at Earth (black cross) is $\left(0 \mathbin{:} 0.21 \mathbin{:} 0.79 \right)_{\oplus}$, the limits are consistent with all compositions possible under averaged oscillations.}
  \label{fig:flav_scan2}
\end{figure}

A related test of the astrophysical flux is to measure the flavor composition of the contained event sample.  Compared with the previous contained event analysis \cite{Aartsen:2015ivb}, this analysis benefits from much better track energy resolution and also the presence of the inelasticity distribution; the inelasticity distribution has some sensitivity to the presence of $\nu_\tau$, since $\nu_\tau$ interactions, followed by $\tau\rightarrow\mu\nu_\mu\overline\nu_\tau$ decays, will lead to events with larger visible inelasticity than $\nu_{\mu}$ interactions of the same energy.     A global fit combining results from contained events and through-going muons found tighter limits that enabled constraints on the source flavor composition, however it compared cascades and tracks in different energy ranges where the energy spectrum may differ \cite{Aartsen:2015knd}.

Figure \ref{fig:flav_scan2} shows confidence levels for various $\nu_e:\nu_\mu:\nu_\tau$ ratios obtained by fitting the data with the same parameters in Tab.~\ref{tab:best_fit_det} as nuisance parameters.  The lines and points show the expectation from different production models and standard neutrino oscillations, including conventional pion decay with source flavor composition $\left(\frac{1}{3}:\frac{2}{3}:0\right)_S$, neutron decay with $\left(1:0:0\right)_S$, and a model where muons lose their energy via synchrotron radiation before they decay with $\left(0:1:0\right)_S$ \cite{Kashti:2005qa}.  All of these conventional scenarios are along a narrow line, and, unfortunately, all are within the 68\% confidence range for the analysis.  The best-fit composition, $\left(0:0.21:0.79\right)_\oplus$, is on the left side of the triangle.   However, because most $\nu_\tau$ produce cascades, there is a relatively high degeneracy between $\nu_\tau$ and $\nu_e$, so the confidence levels nearer the middle of the triangle are high; the break in the degeneracy from the inelasticity distribution of starting tracks is inadequate to statistically separate the $\nu_\tau$ and $\nu_e$ components.   In this analysis, astrophysical cascades and tracks have a similar energy range, with 68\% central ranges of 11 TeV to 410 TeV and 8.6 TeV to 207 TeV, respectively.   This is the tightest limit using samples with a similar energy range; a previous global fit \cite{Aartsen:2015knd} used contained cascades and through-going muons; the latter had a much higher energy range for through-going tracks (330 TeV to 1.4 PeV) than for cascades, where energies above 30 TeV were probed \cite{Aartsen:2015rwa}.   With track and cascade samples having different energies, if the astrophysical spectrum is not a perfect power law, then the confidence regions on the flavor triangle will be shifted.

The extreme compositions of 100\% $\nu_e$ and 100\% $\nu_\mu$ are ruled out with high confidence,  at 5.8$\sigma$ and 7.4$\sigma$, respectively.  In fact, any composition with more than 2/3 $\nu_\mu$  is ruled out at 95\% confidence level.  These constraints can be used to put limits on exotic models.

\subsection{Atmospheric neutrino/antineutrino ratio}

\begin{figure}
  \centering
  \includegraphics[width=0.98\linewidth]{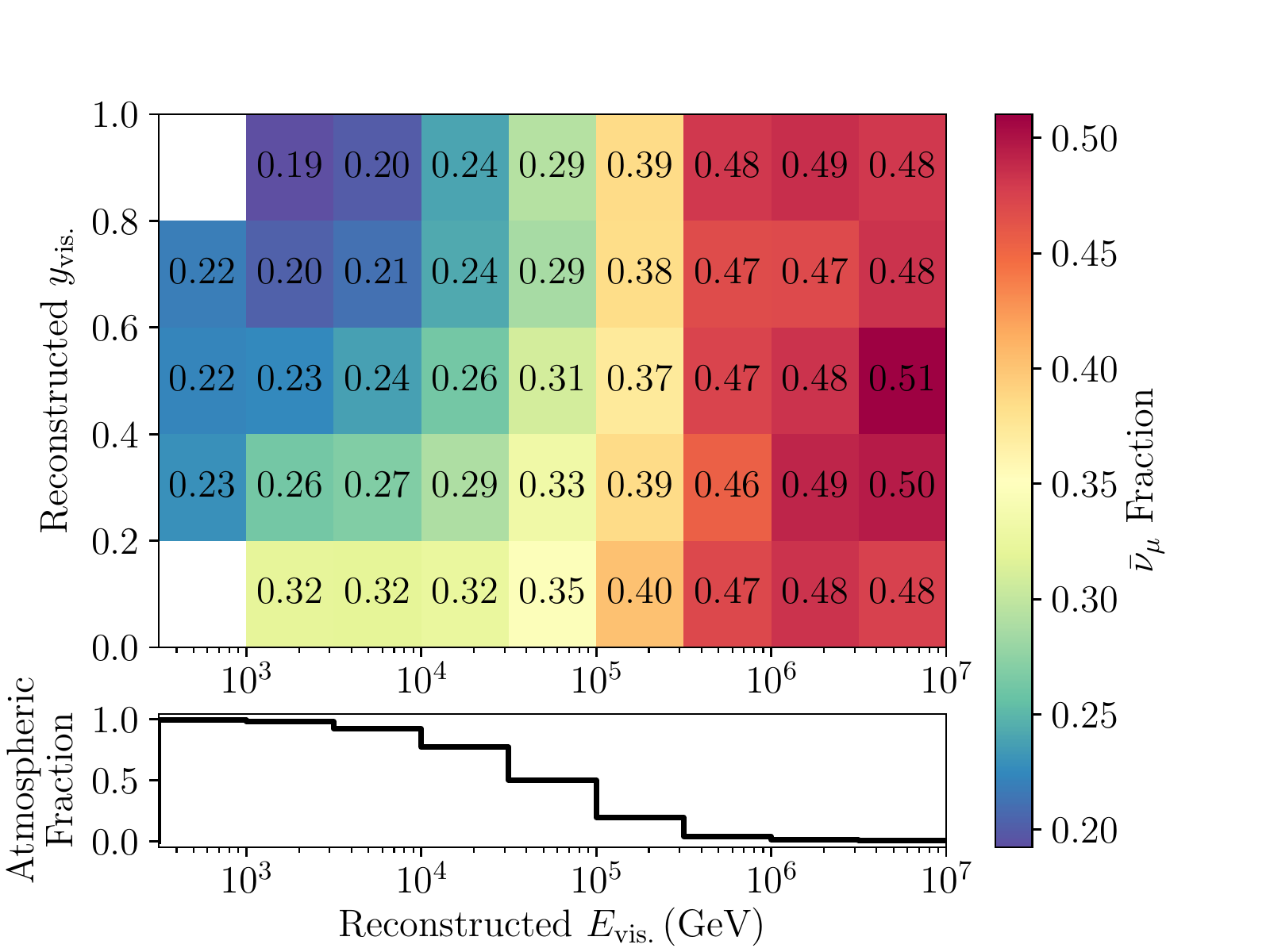}
  \caption{The predicted fraction of $\bar{\nu}_{\mu}$ contributing to the total $\nu_{\mu}+\bar{\nu}_{\mu}$ event rate in bins of reconstructed energy and inelasticity.  At energies below $\sim 10\,\mathrm{TeV}$, there are substantial differences in the inelasticity distribution that enable the atmospheric neutrino to antineutrino ratio to be measured.  The bottom panel shows the fraction of atmospheric neutrinos contributing to the total event rate in bins of reconstructed energy.  At energies above $\sim 100\,\mathrm{TeV}$ where astrophysical neutrinos begin to dominate the event rate, differences in the inelasticity distribution vanish, and it is not possible to measure the neutrino to antineutrino ratio for the astrophysical flux.  An equal neutrino and antineutrino composition is assumed for the astrophysical flux here.}
  \label{fig:inel_nunubar}
\end{figure}

At energies below about 10 TeV, neutrinos and antineutrinos have substantial differences in their inelasticity distributions, since neutrino interactions are more sensitive to quarks,  while antineutrinos are more attuned to antiquarks.  At large Bjorken$-x$ values where valence quarks dominate, the differences are substantial, leading to roughly a factor of 2 difference in cross section \cite{Paukkunen:2010hb,Formaggio:2013kya} as well as a difference in inelasticity distribution.  Unfortunately, as Fig. \ref{fig:inel_nunubar} shows, the difference slowly disappears above 10 TeV.  So, there is little sensitivity to the $\nu_{\mu}:\bar{\nu}_{\mu}$ ratio of astrophysical neutrinos.  However, inelasticity can be used to measure the $\nu_{\mu}:\bar{\nu_{\mu}}$ ratio at lower energies for atmospheric neutrinos.  The atmospheric $\nu_{\mu}:\bar{\nu}_{\mu}$ flux ratio varies with both energy and zenith angle, and we choose to measure an overall scaling factor of the $\nu_{\mu}:\bar{\nu}_{\mu}$ flux ratio from the HKKMS calculation, $R_{\nu_\mu/\overline\nu_\mu}$.  At 1 TeV, the direction-averaged $\nu_{\mu}:\bar{\nu}_{\mu}$ flux ratio is 1.55 and rises slowly to an asymptoic value of 1.75 above 100 TeV.

When the parameter $R_{\nu_\mu/\overline\nu_\mu}$ is added to the list of parameters in Tab.~\ref{tab:best_fit_det}, the best fit value is $R_{\nu_\mu/\overline\nu_\mu} = 0.77^{+0.44}_{-0.25}$.  A flux of 100\% neutrinos (no $\overline\nu$) is disfavored at 3.8$\sigma$, while the reverse is excluded at $5.4\sigma$.  It should be noted that these limits are also dependent on the calculated angular distributions of $\nu$ and $\overline\nu$ in addition to inelasticity.    The sensitive range for this analysis is  770 GeV to 21 TeV; at higher energies, there is little $\nu:\overline\nu$ discrimination.  This is the first $\nu:\overline\nu$ measurement in this energy range.  Along with measurements of the atmospheric muon charge ratio \cite{Agafonova:2014mzx,Khachatryan:2010mw,Adamson:2007ww,Singh:2017orq}, it is a useful diagnostic of particle production in cosmic-ray interactions \cite{Gaisser:2011cc}. However, since this is an overall scaling factor, the same for all energies and zenith angles, it should not be directly interpreted in terms of hadronic interaction models since they also change the energy and zenith distribution assumed.

\subsection{Neutrino charm production}

Inelasticity measurements can also be used to probe charm production in neutrino interactions.   The fraction of CC neutrino interactions that produce charm quarks rises slowly, from 10\% at 100 GeV to 20\% at 100 TeV.  Charm quarks are produced primarily when a neutrino interacts with a strange sea quark; these sea quarks have lower mean Bjorken$-x$ values than valence quarks, so the interactions have flatter inelasticity distributions than neutrino interactions with valence quarks.  There is also a roughly 10\% chance for a charm quark to decay semileptonically and produce an extra muon.   These muons will not be distinguishable from the primary muon from the CC interaction; the energy loss from the two muons will add, and they will be reconstructed as a single, higher-energy muon, leading to a lower apparent inelasticity.   For a $\nu_e$ or $\nu_\tau$ interaction, the presence of a muon from semileptonic charm decay may lead to a track event with an apparent high inelasticity.

\begin{figure}
  \centering
  \includegraphics[width=\linewidth]{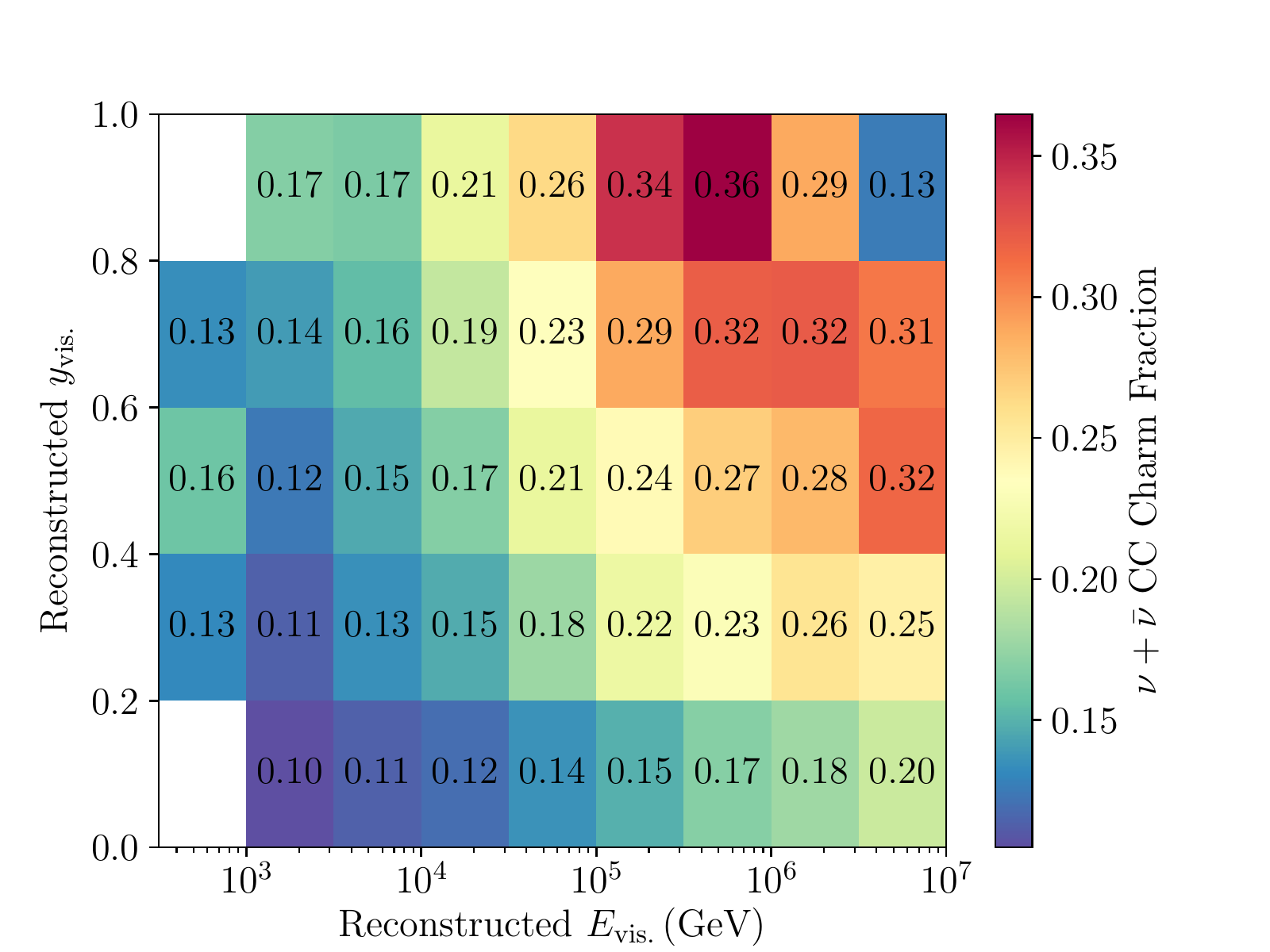}
  \caption{The predicted fractional contribution of all-flavor neutrino CC charm production events to the total event rate in bins of reconstructed visible energy and inelasticity.  The increased charm production fraction at high visible inelasticity and high energy (up to $36\%$) provides a shape difference that allows the presence of charm production to be identified in a likelihood fit to the data.}
  \label{fig:inel_charm}
\end{figure}

The contribution of charm production to different inelasticity events at different energies is shown in Fig. \ref{fig:inel_charm}.    Charm is most visible at energies above 100 TeV and at high inelasticity. For energies between 100 TeV and 1 PeV, more than 1/3 of the events with reconstructed $y_{\rm vis.}>$ 0.8 produce charm.  This shape difference can be used to search for charm production in a maximum likelhood fit.
 
An additional parameter, $R_{\mathrm{CC,charm}}$, scaling the CC charm production event rate is added to the parameters in Tab.~\ref{tab:best_fit_det}.  Fitting cascades and tracks jointly, we find the $68\%$ interval, $R_{\mathrm{CC,charm}} = 0.93^{+0.73}_{-0.59}$.  The test statistic for a null hypothesis with zero charm production is $-2\Delta\ln(L)=2.8$, so zero charm production is excluded at 91\% confidence level.  The 90\% upper limit is 2.3 times the leading-order HERAPDF1.5 prediction.  

The central 90\% of neutrino energies contributing to this test statistic is from 1.5 to 340 TeV.  This is a wider energy range than the central 90\% of charm production events, 1.3 TeV to 44 TeV, because charm production is larger at higher energy.   This is the highest energy measurement of charm production yet, and the upper end of the energy range is above the critical energies of charm hadrons where interactions in ice must occur.   Similar methods could be used to search for other special types of neutrino interactions beyond charm production, including those beyond the Standard Model.
  
\section{Conclusions}

We have developed a tool to measure neutrino inelasticity in Gigaton-scale H$_2$O based detectors and presented the first measurements of neutrino inelasticity in very high-energy (above 1 TeV) neutrino interactions, using a sample of starting track events collected by the IceCube Neutrino Observatory.   The measured inelasticity distributions are in good agreement with the predictions of a modern NLO calculation.  More data is needed to reach anticipated theoretical uncertainties in these calculations.

We have made a global fit to these neutrino data, fitting cascades in two dimensions: energy and zenith angle, and starting tracks in three dimensions: energy, zenith angle and inelasticity, to extract information about the astrophysical and atmospheric neutrino fluxes.   This fit finds an astrophysical power-law spectral index of $\gamma=2.62\pm 0.07$, in good agreement with previous fits to contained events and cascades, but in tension with previous results based on through-going muons, a sample that is generally higher in energy than the contained event samples.   To explore this tension, we performed a fit where we allowed the astrophysical flux to float separately for cascades and starting tracks, with different spectral indices.  Unfortunately, this leads to a spectral index for the tracks, $\gamma=2.43^{+0.28}_{-0.30}$, intermediate between the combined result and that for through-going tracks, with an error that is consistent with either.  

We then relaxed the requirement that the astrophysical neutrino flavor ratio be $\left(\frac{1}{3}:\frac{1}{3}:\frac{1}{3}\right)_{\oplus}$, and calculated the confidence level for other compositions.  We found a best fit point consisting of 79\% $\nu_\tau$, 21\% $\nu_\mu$, but with a broad allowed contour that encompasses all of the models that invoke conventional acceleration mechanisms and standard neutrino oscillations.  More exotic models may be ruled out.

We also set limits on the $\nu_\mu:\bar{\nu}_\mu$ ratio in atmospheric neutrinos and exclude zero production of charm quarks in neutrino interactions at 91\% confidence level.  This is only the second study, after measurements of the cross section using neutrinos with energies above 1 TeV.  

Using the indirect signature in the inelasticity distribution, we observe, at greater than 90\% confidence level, CC charm production in neutrino interactions, at energies between 1.5 and 340 TeV, more than an order of magnitude higher in energy than accelerator measurements.

Looking ahead, we expect that IceCube-Gen2 \cite{Aartsen:2017pja} and KM3NeT2.0 \cite{Adrian-Martinez:2016fdl} will collect larger samples of contained events, which can be used to make more precise measurements of inelasticity.   These detectors could collect substantial samples of events with energies above 100 TeV.   With the increased precision, it will also be possible to study several new topics.  Tau neutrinos are one example; $\nu_\tau$ interactions have a distinctive inelasticity distribution which could be used to detect a $\nu_\tau$ signal.  Top quark production may also be accessible if enough energetic neutrinos are available.  One calculation found that, for 10 PeV neutrinos, top quarks are produced in 5\% of the interactions  \cite{Barge:2016uzn}.  It may also be possible to study other Standard Model neutrino interactions, such as diffractive production of W bosons in the Coulomb field of oxygen nuclei  \cite{Seckel:1997kk,Alikhanov:2014uja}; the cross section for $\nu + O \rightarrow  l + W^+ + X$ is about 8\% of the charged-current cross section for 1 PeV neutrinos.  Even the first phase of IceCube-Gen2 should enable improved calibrations of the existing data, reducing the systematic uncertainties.  With moderately improved calibrations, the precision of the inelasticity measurements should scale as the square root of the effective volume times the live time. 

With an improved surface veto to reject atmospheric neutrinos, it might also be possible to measure the $\nu:\overline\nu$ ratio of astrophysical neutrinos.   If one could use the self-veto and a surface air-shower-array veto to reject atmospheric neutrinos with energies in the 1-10 TeV energy range, the inelasticity distribution could be used to determine the $\nu:\overline\nu$  ratio of astrophysical neutrinos.

The data could also be used to search for beyond-standard-model (BSM) physics, such as supersymmetry \cite{Carena1998}, leptoquarks \cite{Anchordoqui2006} or quantum gravity with a relatively low scale \cite{Anchordoqui2007}.  These phenomena also produce cross section enhancements which could be visible via increased neutrino absorption in the Earth, but the inelasticity distribution has a higher diagnostic utility than a simple increase in neutrino absorption.   The use of inelasticity allows for a more sensitive search than by merely counting cascades and tracks \cite{Becirevic:2018uab}.
A combined fit to cross section and inelasticity measurements would provide even better constraints on new physics.

For most of these phenomena, the LHC provides better limits compared to IceCube Gen2 and KM3NeT 2.0.  However, experiments that aim to record the coherent radio Cherenkov emission from ultra-energetic neutrinos with energies above $10^{17}$ eV can reach supra-LHC energies.  The ARA \cite{Allison:2015eky} and ARIANNA \cite{Barwick:2014pca} collaborations both propose to deploy large ($> 100$ km$^3$) arrays that will, unless ultra-high energy cosmic-rays are primarily iron,  collect useful (order 100 events) samples of cosmogenic neutrinos. The challenge here is that these experiments are primarily sensitive to cascades, while the energy deposition from tracks is too diffuse to be observable.  However, it may be possible to take advantage of the Landau-Pomeranchuk-Migdal (LPM) effect to separate electromagnetic showers, which at energies above $10^{20}$ eV are elongated, from the hadronic showers from the target nucleus, which are less subject to the LPM effect.   This leads to a moderately elongated electromagnetic shower following a compact hadronic shower \cite{Gerhardt:2010bj}.  With this, it might be possible to measure the inelasticity of charged-current $\nu_e$ interactions \cite{Alvarez1999}. 


\begin{acknowledgments}
  The IceCube Collaboration designed, constructed and
  now operates the IceCube Neutrino Observatory. Data
  processing and calibration, Monte Carlo simulations of
  the detector and of theoretical models, and data analyses
  were performed by a large number of collaboration members,
  who also discussed and approved the scientific results
  presented here. The main authors of this manuscript
  were Gary Binder and Spencer Klein. It was reviewed by the entire collaboration
  before publication, and all authors approved
  the final version of the manuscript.
  
  We acknowledge the support from the following agencies:
  USA -- U.S. National Science Foundation-Office of Polar Programs,
  U.S. National Science Foundation-Physics Division,
  Wisconsin Alumni Research Foundation,
  Center for High Throughput Computing (CHTC) at the University of Wisconsin-Madison,
  Open Science Grid (OSG),
  Extreme Science and Engineering Discovery Environment (XSEDE),
  U.S. Department of Energy-National Energy Research Scientific Computing Center,
  Particle astrophysics research computing center at the University of Maryland,
  Institute for Cyber-Enabled Research at Michigan State University,
  and Astroparticle physics computational facility at Marquette University;
  Belgium -- Funds for Scientific Research (FRS-FNRS and FWO),
  FWO Odysseus and Big Science programmes,
  and Belgian Federal Science Policy Office (Belspo);
  Germany -- Bundesministerium f\"ur Bildung und Forschung (BMBF),
  Deutsche Forschungsgemeinschaft (DFG),
  Helmholtz Alliance for Astroparticle Physics (HAP),
  Initiative and Networking Fund of the Helmholtz Association,
  Deutsches Elektronen Synchrotron (DESY),
  and High Performance Computing cluster of the RWTH Aachen;
  Sweden -- Swedish Research Council,
  Swedish Polar Research Secretariat,
  Swedish National Infrastructure for Computing (SNIC),
  and Knut and Alice Wallenberg Foundation;
  Australia -- Australian Research Council;
  Canada -- Natural Sciences and Engineering Research Council of Canada,
  Calcul Qu\'ebec, Compute Ontario, Canada Foundation for Innovation, WestGrid, and Compute Canada;
  Denmark -- Villum Fonden, Danish National Research Foundation (DNRF);
  New Zealand -- Marsden Fund;
  Japan -- Japan Society for Promotion of Science (JSPS)
  and Institute for Global Prominent Research (IGPR) of Chiba University;
  Korea -- National Research Foundation of Korea (NRF);
  Switzerland -- Swiss National Science Foundation (SNSF).

\end{acknowledgments}


\begin{thebibliography}{99}

  \bibitem{Aartsen:2014gkd} 
  M.~G.~Aartsen {\it et al.} [IceCube Collaboration],
  Phys.\ Rev.\ Lett.\  {\bf 113}, 101101 (2014).
  
  \bibitem{Aartsen:2013jdh} 
  M.~G.~Aartsen {\it et al.} [IceCube Collaboration],
  Science {\bf 342}, 1242856 (2013).

\bibitem{Aartsen:2015zva}
  L.~Mohrmann [IceCube Collaboration],
  PoS ICRC {\bf 2015} (2016) 1066.
  
  
  
\bibitem{Aartsen:2017mau}
  H.~M.~Niederhausen and Y.~Xu [IceCube Collaboration],
  PoS ICRC {\bf 2017} (2018) 968.
  
\bibitem{Aartsen:2015ivb} 
  M.~G.~Aartsen {\it et al.} [IceCube Collaboration],
  Phys.\ Rev.\ Lett.\  {\bf 114}, no. 17, 171102 (2015).
    
\bibitem{Aartsen:2015knd} 
  M.~G.~Aartsen {\it et al.} [IceCube Collaboration],
  Astrophys.\ J.\  {\bf 809}, no. 1, 98 (2015).  
\bibitem{Aartsen:2017kpd}
  M.~G.~Aartsen {\it et al.} [IceCube Collaboration],
    Nature {\bf 551}, 596 (2017).
  
\bibitem{Dissertation}G.~Binder, PhD dissertation, University of California, Berkeley, (2017).  Available at \url{https://docushare.icecube.wisc.edu/dsweb/Get/Document-82240/thesis.pdf} 

 \bibitem{Schienbein:2007fs} 
  I.~Schienbein, J.~Y.~Yu, C.~Keppel, J.~G.~Morfin, F.~Olness and J.~F.~Owens,
  Phys.\ Rev.\ D {\bf 77}, 054013 (2008).

 \bibitem{Aubert:1974en} 
  B.~Aubert {\it et al.},
  Phys.\ Rev.\ Lett.\  {\bf 33}, 984 (1974).
  
\bibitem{DIS} R.~Devinish and A.~Cooper-Sarkar,
{\it Deep Inelastic Scattering}, Oxford University Press, 2004.
    
  \bibitem{Kalantarians:2017mkj} 
  N.~Kalantarians, C.~Keppel and M.~E.~Christy,
  Phys.\ Rev.\ C {\bf 96}, no. 3, 032201 (2017).
  
 \bibitem{CooperSarkar:2011pa} 
  A.~Cooper-Sarkar, P.~Mertsch and S.~Sarkar,
  JHEP {\bf 1108}, 042 (2011). 
  
\bibitem{Radescu2013}
  V.~Radescu, [H1 and ZEUS Collaborations], Proceedings, 35th Int. Conf. High energy Phys. (ICHEP 2010) Paris, Fr. July 22-28, 2010.
  
\bibitem{Lai:2010vv} 
  H.~L.~Lai, M.~Guzzi, J.~Huston, Z.~Li, P.~M.~Nadolsky, J.~Pumplin and C.-P.~Yuan,
  Phys.\ Rev.\ D {\bf 82}, 074024 (2010).
  
  \bibitem{Connolly:2011vc} 
  A.~Connolly, R.~S.~Thorne and D.~Waters,
  Phys.\ Rev.\ D {\bf 83}, 113009 (2011).
  
  \bibitem{Barge:2016uzn} 
  V.~Barger, E.~Basso, Y.~Gao and W.~Y.~Keung,
  Phys.\ Rev.\ D {\bf 95}, no. 9, 093002 (2017).
 
 \bibitem{charm}A. Bueno and A. Gascón, Comput. Phys. Commun. {\bf 185}, 638 (2014).
 
  \bibitem{Mousseau:2016snl} 
  J.~Mousseau {\it et al.} [MINERvA Collaboration],
  Phys.\ Rev.\ D {\bf 93}, no. 7, 071101 (2016).  

  
 \bibitem{Schienbein:2009kk} 
  I.~Schienbein, J.~Y.~Yu, K.~Kovarik, C.~Keppel, J.~G.~Morfin, F.~Olness and J.~F.~Owens,
  Phys.\ Rev.\ D {\bf 80}, 094004 (2009). 

 \bibitem{Seckel:1997kk} 
  D.~Seckel,
  Phys.\ Rev.\ Lett.\  {\bf 80}, 900 (1998).

  \bibitem{Alikhanov:2014uja} 
  I.~Alikhanov,
  Phys.\ Lett.\ B {\bf 741}, 295 (2015).
  
  \bibitem{Honda:2006qj} 
  M.~Honda, T.~Kajita, K.~Kasahara, S.~Midorikawa and T.~Sanuki,
  Phys.\ Rev.\ D {\bf 75}, 043006 (2007).
  
 \bibitem{Aartsen:2016xlq} 
  M.~G.~Aartsen {\it et al.} [IceCube Collaboration],
  Astrophys.\ J.\  {\bf 833}, 3 (2016).
  
\bibitem{Aartsen:2015xup} 
  M.~G.~Aartsen {\it et al.} [IceCube Collaboration],
  Phys.\ Rev.\ D {\bf 91}, 122004 (2015).
  
 \bibitem{Aartsen:2017nbu} 
  M.~G.~Aartsen {\it et al.} [IceCube Collaboration],
  Eur.\ Phys.\ J.\ C {\bf 77},  692 (2017).
  
  \bibitem{Aartsen:2012uu} 
  M.~G.~Aartsen {\it et al.} [IceCube Collaboration],
  Phys.\ Rev.\ Lett.\  {\bf 110}, no. 15, 151105 (2013).
  
  \bibitem{Bhattacharya:2015jpa} 
  A.~Bhattacharya, R.~Enberg, M.~H.~Reno, I.~Sarcevic and A.~Stasto,
  JHEP {\bf 1506}, 110 (2015).
  
\bibitem{Garzelli2015}
  M.~V.~Garzelli, S.~Moch and G.~Sigl,
  JHEP 1510, 115 (2015). 

\bibitem{Gauld2016}
  R.~Gauld, J.~Rojo, L.~Rottoli, S.~Sarkar and J.~Talbert,
  JHEP 1602, 130 (2016).
  
  \bibitem{Gaisser:2014bja} 
  T.~K.~Gaisser, K.~Jero, A.~Karle and J.~van Santen,
  Phys.\ Rev.\ D {\bf 90}, no. 2, 023009 (2014).

\bibitem{CORSIKA}D. Heck {\it  et al.}, CORSIKA a Monte Carlo code to simulate extensive air showers., T.. Forschungszentrum Karlsruhe GmbH, Karlsruhe (Germany)., Feb 1998, V + 90 p., TIB Hann. D-30167 Hann. (Germany). (1998).
  
\bibitem{PREM}A. M. Dziewonski and D. L. Anderson, Phys. Earth Planetary Interiors {\bf 25}, 297 (1981). 

\bibitem{Radel:2012ij} 
  L.~Radel and C.~Wiebusch,
  Astropart.\ Phys.\  {\bf 44}, 102 (2013).

 \bibitem{Klein:1998du} 
  S.~Klein,
  Rev.\ Mod.\ Phys.\  {\bf 71}, 1501 (1999).
  
  \bibitem{Chirkin:2004hz} 
  D.~Chirkin and W.~Rhode,
  hep-ph/0407075.
 
\bibitem{Koehne:2013gpa} 
  J.~H.~Koehne, K.~Frantzen, M.~Schmitz, T.~Fuchs, W.~Rhode, D.~Chirkin and J.~Becker Tjus,
  Comput.\ Phys.\ Commun.\  {\bf 184}, 2070 (2013).
  
\bibitem{DeYoung2007}
  T.~DeYoung, S.~Razzaque, and D.~Cowen,
  Astropart. Phys. {\bf 27}, 238 (2007).

\bibitem{Lellis:2002}
  G.~De Lellis, F.~Di Capua, and P.~Migliozzi,
  Phys.\ Lett.\ B 550, 16 (2002).

\bibitem{genie}
  C.~Andreopoulous {\it et al.},
  arXiv:1510.05494
 
   \bibitem{Aartsen:2013vja} 
  M.~G.~Aartsen {\it et al.} [IceCube Collaboration],
  JINST {\bf 9}, P03009 (2014)
  doi:10.1088/1748-0221/9/03/P03009.

 \bibitem{Aartsen:2013ola} 
   D. Chirkin [IceCube Collaboration], in {\it Proc 33rd
International Cosmic Ray Conference (ICRC2013): Rio
de Janeiro, Brazil, 2013}, p. 0580.
  
  \bibitem{Aartsen:2013rt} 
  M.~G.~Aartsen {\it et al.} [IceCube Collaboration],
  Nucl.\ Instrum.\ Meth.\ A {\bf 711}, 73 (2013).

  \bibitem{Rongen:2016sbk} 
  M.~Rongen [IceCube Collaboration],
  EPJ Web Conf.\  {\bf 116}, 06011 (2016).
    
\bibitem{Halzen:2010yj} 
  F.~Halzen and S.~R.~Klein,
  Rev.\ Sci.\ Instrum.\  {\bf 81}, 081101 (2010).
  
  \bibitem{Aartsen:2016nxy} 
  M.~G.~Aartsen {\it et al.} [IceCube Collaboration],
  JINST {\bf 12}, no. 03, P03012 (2017).
  
 \bibitem{Collaboration:2011ym} 
  R.~Abbasi {\it et al.} [IceCube Collaboration],
  Astropart.\ Phys.\  {\bf 35}, 615 (2012).

  \bibitem{Achterberg:2006md} 
  A.~Achterberg {\it et al.} [IceCube Collaboration],
  Astropart.\ Phys.\  {\bf 26}, 155 (2006).
  
  \bibitem{Abbasi:2008aa} 
  R.~Abbasi {\it et al.} [IceCube Collaboration],
  Nucl.\ Instrum.\ Meth.\ A {\bf 601}, 294 (2009).
  
  \bibitem{Abbasi:2010vc} 
  R.~Abbasi {\it et al.} [IceCube Collaboration],
  Nucl.\ Instrum.\ Meth.\ A {\bf 618}, 139 (2010)
  
  \bibitem{Aartsen:2014muf} 
  M.~G.~Aartsen {\it et al.} [IceCube Collaboration],
  Phys.\ Rev.\ D {\bf 91},  022001 (2015).
  
 \bibitem{Jakob}J. van Santen, Ph. D. thesis, University of Wisconsin, Madison (2014).
  
 \bibitem{Ahn2009}
  E.~Ahn {\it et al.}
  Phys.\ Rev.\ D {\bf{80}}, 094003 (2009).
 
\bibitem{Gaisser:2011cc} 
  T.~K.~Gaisser,
  Astropart.\ Phys.\  {\bf 35}, 801 (2012).
 
 \bibitem{Ahrens:2003ix} 
  J.~Ahrens {\it et al.} [IceCube Collaboration],
  Astropart.\ Phys.\  {\bf 20}, 507 (2004).

\bibitem{Abbasi:2012wht} 
  R.~Abbasi {\it et al.} [IceCube Collaboration],
  Nucl.\ Instrum.\ Meth.\ A {\bf 703}, 190 (2013).

\bibitem{FC}
  G.~J.~Feldman and R.~D.~Cousins,
  Phys.\ Rev.\ D {\bf 57}, 3873 (1998).
  

\bibitem{Pena2001}
  J.A.~Castro Pena, G.~Parente, E.~Zas,
  Phys.\ Lett.\ B, {\bf 50}, 1–2, 125-132, (2001). 

\bibitem{PDG}
M.~Tanabashi {\it et al.} (Particle Data Group),
Phys.\ Rev.\ D {\bf 98}, 030001 (2018).
  \bibitem{Enberg:2008te} 
  R.~Enberg, M.~H.~Reno and I.~Sarcevic,
  Phys.\ Rev.\ D {\bf 78}, 043005 (2008)
  doi:10.1103/PhysRevD.78.043005
  [arXiv:0806.0418 [hep-ph]].
 
\bibitem{Esteban2017}I. Esteban, M. C. Gonzalez-Garcia, M. Maltoni, et al., J. High Energy Phys. 2017, 87 (2017).     
      
\bibitem{Kashti:2005qa} 
  T.~Kashti and E.~Waxman,
  Phys.\ Rev.\ Lett.\  {\bf 95}, 181101 (2005).
  
\bibitem{Aartsen:2015rwa} 
  M.~G.~Aartsen {\it et al.} [IceCube Collaboration],
  Phys.\ Rev.\ Lett.\  {\bf 115}, no. 8, 081102 (2015)
  doi:10.1103/PhysRevLett.115.081102.
  
\bibitem{Paukkunen:2010hb} 
  H.~Paukkunen and C.~A.~Salgado,
  JHEP {\bf 1007}, 032 (2010)
  doi:10.1007/JHEP07(2010)032
  [arXiv:1004.3140 [hep-ph]].
  
\bibitem{Formaggio:2013kya} 
  J.~A.~Formaggio and G.~P.~Zeller,
  Rev.\ Mod.\ Phys.\  {\bf 84}, 1307 (2012).  
  
 \bibitem{Agafonova:2014mzx} 
  N.~Agafonova {\it et al.} [OPERA Collaboration],
  Eur.\ Phys.\ J.\ C {\bf 74}, 2933 (2014).
  
  \bibitem{Khachatryan:2010mw} 
  V.~Khachatryan {\it et al.} [CMS Collaboration],
  Phys.\ Lett.\ B {\bf 692}, 83 (2010)
  doi:10.1016/j.physletb.2010.07.033
  [arXiv:1005.5332 [hep-ex]].
  
\bibitem{Adamson:2007ww} 
  P.~Adamson {\it et al.} [MINOS Collaboration],
  Phys.\ Rev.\ D {\bf 76}, 052003 (2007).

\bibitem{Singh:2017orq} 
  J.~Singh and J.~Singh,
  arXiv:1709.01064 [physics.ins-det].
  
 \bibitem{Aartsen:2017pja} 
  M.~Aartsen {\it et al.} [IceCube Gen2 Collaboration],
  arXiv:1710.01207 [astro-ph.IM].
    
  \bibitem{Adrian-Martinez:2016fdl} 
  S.~Adrian-Martinez {\it et al.} [KM3Net Collaboration],
  J.\ Phys.\ G {\bf 43}, no. 8, 084001 (2016).
  
\bibitem{Becirevic:2018uab} 
  D.~Be\u{c}irevi\'{c}, B.~Panes, O.~Sumensari and R.~Zukanovich Funchal,
  arXiv:1803.10112 [hep-ph].
  
  \bibitem{Anchordoqui2006}
  L.~A.~Anchordoqui, C.~A.~García Canal, H.~Goldberg, {\it et al.},
  Phys.\ Rev.\ D {\bf 74}, 125021 (2006).

  \bibitem{Anchordoqui2007}
  L.~A. Anchordoqui, M.~M. Glenz, and L.~Parker,
  Phys.\ Rev.\ D {\bf 75}, 024011 (2007).

  \bibitem{Carena1998}
  M.~Carena, D.~Choudhury, S.~Lola, et al.,
  Phys.\ Rev.\ D {\bf 58}, 095003 (1998).

  \bibitem{Allison:2015eky} 
  P.~Allison {\it et al.} [ARA Collaboration],
  Phys.\ Rev.\ D {\bf 93}, no. 8, 082003 (2016).
  
  \bibitem{Barwick:2014pca} 
  S.~W.~Barwick {\it et al.} [ARIANNA Collaboration],
  Astropart.\ Phys.\  {\bf 70}, 12 (2015).
  
  \bibitem{Gerhardt:2010bj} 
  L.~Gerhardt and S.~R.~Klein,
  Phys.\ Rev.\ D {\bf 82}, 074017 (2010).

  \bibitem{Alvarez1999}
  J.~Alvarez-Muniz, R.~A.~Vazquez, and E.~Zas,
  Phys.\ Rev.\ D {\bf 61}, 023001 (1999).

  

  
  
    
  


  
    
  
  
  
  
  
 

  


  
  
  
  
  
  
  


  
  


  
 
  


 
 

  

    
  
  

  
  
  
  
  

 

 
  

  
      
  
  
  
  

  
 
    
  
  
  
  






  

\end{thebibliography}
\end{document}